\documentclass[aps,twocolumn]{revtex4} 
\usepackage{graphicx}
\usepackage[centertags]{amsmath} 
\usepackage{amssymb}

\newcommand{\nn}{\nonumber} 
  
\newcommand{\bb}[1]{\mathbb{#1}}
\newcommand{\bphi}{\boldsymbol{\varphi}}
\newcommand{\bsigma}{\boldsymbol{\sigma}}

\newcommand{\bnabla}{\boldsymbol{\nabla}}

\DeclareMathOperator{\e}{e}
\DeclareMathOperator{\grad}{grad}

\begin{document}

\title{Full control of qubit rotations in a voltage-biased
  superconducting flux qubit} 
\author{Luca Chirolli}
\author{Guido Burkard} 

\affiliation{Department of Physics and Astronomy,
  University of Basel, Klingelbergstrasse 82,
  CH-4056 Basel, Switzerland}

\begin{abstract}
We study a voltage-controlled version of the superconducting flux
qubit [Chiorescu \textit{et al.}, Science {\bf 299}, 1869 (2003)]
and show that full control of qubit rotations on the entire Bloch 
sphere can be achieved.
Circuit graph theory is used to study a setup where
voltage sources are attached to the two superconducting islands
formed between the three Josephson junctions in the flux qubit.
Applying a voltage allows qubit rotations about the $y$ axis,
in addition to pure $x$ and $z$ rotations obtained in the absence of 
applied voltages.
The orientation and magnitude of the rotation axis on the Bloch 
sphere can be tuned by the gate voltages, the external magnetic flux,
and the ratio $\alpha$ between the Josephson energies of the junctions
via a flux-tunable junction. 
We compare the single-qubit control in the known regime $\alpha<1$
with the unexplored range $\alpha>1$ and estimate the decoherence
due to voltage fluctuations.
\end{abstract}

\maketitle


\section{Introduction}
\label{sec:introduction}

\begin{figure}[t]
 \begin{center}
\includegraphics[width=7cm]{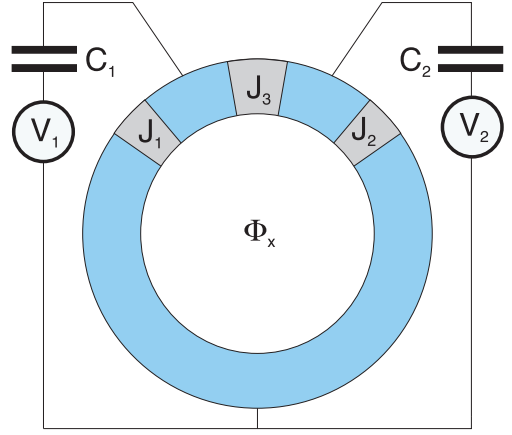}
    \caption{The voltage-biased SC flux qubit (schematic).
    The circuit consists of a SC ring (blue) with three
    Josephson junctions $J_1$, $J_2$, and $J_3$,
    threaded by an external magnetic flux $\Phi_x$.
    The Josephson energy of the middle junction $J_3$
    differs from the other two by a factor of $\alpha$.
    A voltage bias $V_i$ is applied to each of the two
    islands formed by the three junctions via a capacitor $C_i$.
    \label{fig:schematic}}
 \end{center}
\end{figure}

Superconducting (SC) circuits can exhibit a great
variety of quantum mechanical phenomena and are studied for their 
potential as devices for quantum information processing.
Several different circuit implementations of a SC quantum bit
(qubit) have been investigated both theoretically
and experimentally \cite{Makhlin01,Devoret04}.

A prototype of a SC flux qubit, characterized by a working regime in
which the Josephson energy dominates over the charging energy,
$E_J\gg E_C$,  has been theoretically designed and experimentally
realized
\cite{Orlando99,Mooij99,VanDerWal00,Friedman00,Chiorescu03,Bertet05,Yoshihara06}, 
showing quantum
superposition and coherent evolution of two macroscopic  states
carrying opposite persistent currents that represent  the qubit
states.
The flux qubit state is related to a magnetic moment, and is thus
typically controlled via the application of external magnetic fields
which create magnetic flux through the loop(s) in the circuit.
An advantage of flux qubits is their relative insensitivity to charge
fluctuations that can lead to fast decoherence
\cite{LinTian99,LinTian02,vanderWal03}, while magnetic fluctuations are typically 
more benign.

A second type of SC qubits, the so called charge qubits
\cite{Averin98,Makhlin99,Nakamura99,Pashkin03,Yamamoto03}, operates in the
limit in which the charge energy dominates, $E_C\gg E_J$,  thus being
relatively insensitive to magnetic fluctuations, while  having a well
defined value of the charge on a SC island, in  which the presence or
absence of an extra Cooper pair determines  the state of the qubit.
The intermediate regime in which the Josephson and charge energies
are comparable, $E_J\approx E_C$, has been investigated and  realized
in the ``quantronium'' \cite{quantronium}.
Another type of qubit is the Josephson, or phase, qubit, consisting
of a single junction \cite{Martinis02}.

In this paper, we investigate the possibility of enhancing the  control
of a SC flux qubit via the application of electrostatic gates
\cite{Orlando99,Amin,AminR}.  We study the flux qubit proposed by
Orlando \textit{et al.} \cite{Orlando99}. 
While in \cite{Orlando99}, the effect of any applied voltages was kept low
in order to avoid charge noise, we explore the possibility
of making use of the off-set gate charge as an additional control variable.  
We define two device parameters.  Assuming for simplicity two Josephson
junctions to have equal Josephson energies ($E_{J1}=E_{J2}=E_J$), the 
first parameter is given by the ratio $\alpha=E_{J3}/E_J$ between the 
Josephson energy of the third junction and the remaining two junctions.
The regime of interest here is $0.5<\alpha\lesssim 1.5$ although in
principle larger values are possible.
The second parameter is the ratio between
the Josephson energy and the charging energy, $E_J/E_C$ which
for flux qubits is typically about $10$ or larger.
We analyze the role of these parameters in detail and, in addition to
the well-studied regime $\alpha<1$, also explore the opposite regime $\alpha>1$.
Particular effort is spent looking
for a single-qubit Hamiltonian  in which an effective pseudo-magnetic field
couples to all three  components of the pseudo-spin represented by the
circuit.  A charge qubit in which a $\sigma_y$ term in the single-qubit
Hamiltonian has been proposed in \cite{Falci00}.
The possibility of changing the relative phase of the qubit states,
together with the capability to flip them, allows full control over the
qubit.  
Full control on the Bloch sphere is thought to be very
useful in the field of adiabatic quantum computation
\cite{Aharonov,TO,BTD}.

Circuit theory provides us with a systematic and universal 
method for analyzing any electrical circuit that can be
represented by lumped elements \cite{BKD,charge,BDBCM,DBK05}.
Through the language of a graph theoretic
formalism, Kirchhoff's laws and the Hamiltonian of the circuit
are written in terms of a set of independent canonical
coordinates that can easily be quantized.
The formalism of \cite{BKD,charge,BDBCM} is particularly suited for
studying circuits containing superconducting elements, like Josephson
junctions, that are treated as nonlinear inductors.   Here,
we make use of the extended circuit theory that accounts for charging 
effects and can be applied both for charge and flux qubits \cite{charge}. 

Our main result is the identification of the parameter range
for $\alpha$ and $E_J/E_C$ in the voltage-controlled flux qubit
in which the single qubit Hamiltonian acquires a $\sigma_y$ 
term in addition to the $\sigma_x$ and $\sigma_z$ terms, thus 
allowing full control of the qubit rotations on the Bloch sphere.
In this regime, we compute the dependence of the single-qubit 
Hamiltonian on the applied voltages $V_1$ and $V_2$.
For the quantitative analysis of the qubit dynamics we calculate
the tunneling amplitudes appearing in the Hamiltonian as functions of
the device parameters.

The paper is structured as follows. In Section \ref{sec:circuit} we
briefly review circuit theory \cite{BKD,charge,BDBCM,DBK05} 
and apply it to the circuit of Fig.~\ref{fig:circuit} to find its
Hamiltonian.
Section \ref{sec:born-oppenheimer} contains the derivation of
the effective periodic potential in the Born-Oppenheimer approximation. 
In Section \ref{sec:quantum-dynamics}, we address the quantum
dynamics of the circuit and find localized solutions in the
periodic potential.
In Section \ref{sec:bloch-theory} we apply Bloch's theory in a 
tight-binding approximation to find general solutions
in the presence of a voltage bias. 
Sec.~\ref{sec:t1t2} describes the calculation of the tunneling matrix
elements appearing in the qubit Hamiltonian and their dependence
on the device parameters $\alpha$ and $E_J/E_C$. 
In Sec.~\ref{sec:full-control}, we explore the
regime ($\alpha>1$) and show that a full control on the qubit
Hamiltonian is feasible.
In Section \ref{sec:decoherence}, we study the decoherence of the qubit
due to the attached voltage sources.
Finally, Sec.~\ref{sec:results} contains a summary of our results
and conclusions.


\section{The circuit}
\label{sec:circuit}

\begin{figure}[b]
 \begin{center}
  \includegraphics[width=8.5cm]{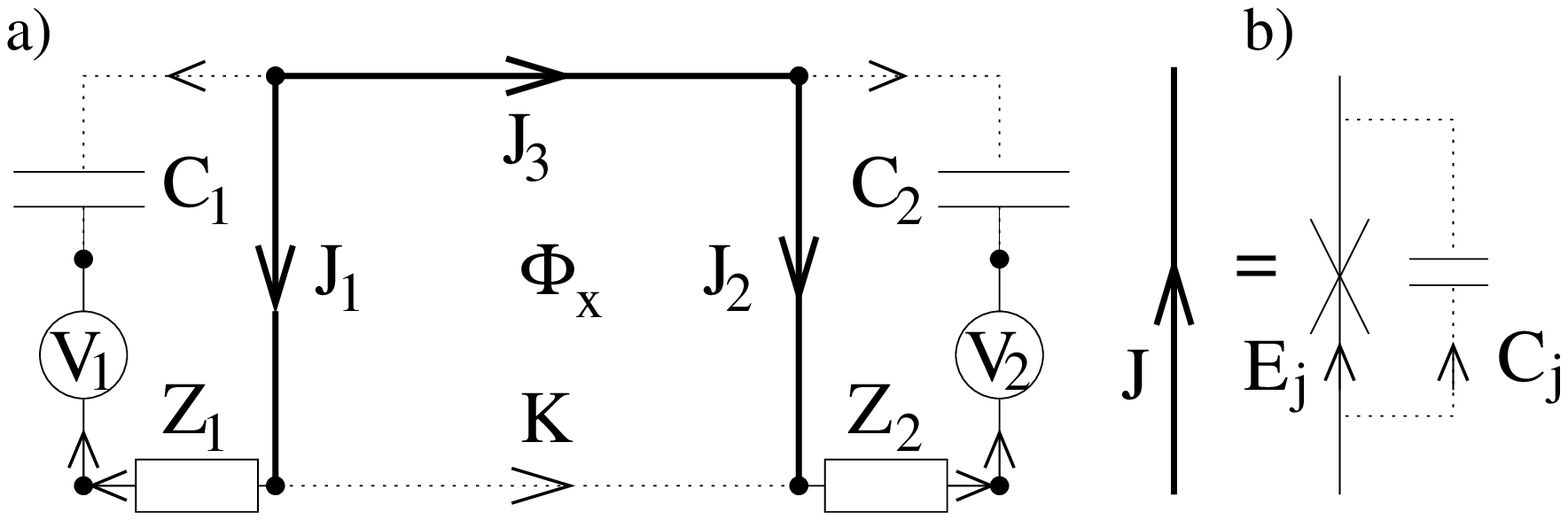}
    \caption{a) Circuit of a voltage-biased flux qubit
             (Fig.~\ref{fig:schematic}). The main loop
             contains three Josephson junctions and a (chord) inductance
             (K). An external magnetic flux $\Phi_x$ threads
             the SC loop. The junctions  $J_1$
             and $J_2$ are biased by two electrostatic gates,
             representing the main new feature of the circuit.  Solid
             lines represent the tree of the circuit graph, while  dotted
             lines are the chords.  b) Each thick solid line represents
             a Josephson junction  shunted by a capacitance $C_J$.
             \label{fig:circuit}}
 \end{center}
\end{figure}
Here we study a version of the Delft flux qubit
\cite{Orlando99,Chiorescu03} with an additional voltage control
(Fig.~\ref{fig:schematic}).
Typically, such a qubit circuit also comprises a readout SQUID which
can be surrounding or attached to the qubit.  We concentrate on the
qubit itself here and do not include the SQUID in our analysis
because the presence of a readout circuit does not alter the analysis
and results for single-qubit control presented here.
A circuit representation of the studied device is shown in
Fig.~\ref{fig:circuit}.
The main loop contains three Josephson
junctions and the loop self-inductance (K),
and is threaded by an external magnetic flux $\Phi_x$.
The junctions form two SC islands to which
electrostatic gates with capacitance $C_1$ and $C_2$ are attached and
voltages  $V_1$ and $V_2$ are applied.  
The voltage sources represent the new elements in the circuit.
As long as the junctions are built in such a way that the Josephson
energy dominates, $E_J\gg E_C$, the qubit is encoded
in the orientation of the circulating persistent current,
as in Refs.~\onlinecite{Orlando99,Chiorescu03}.

We represent the circuit as the oriented graph ${\cal G}$
shown in Fig.~\ref{fig:circuit}a,  consisting of $N=8$ nodes (black dots) 
$n_i$ ($i=1,\ldots,8$) and $B=13$ branches (thin lines)
$b_i$ ($i=1,\ldots,13$), in  which each branch $b_i$ represents one 
of the following lumped  circuit elements:
a (bare) Josephson junction $J$, capacitance $C$,
inductance $K$, voltage source $V$, and impedance $Z$.\
The impedances $Z_1$ and $Z_2$ model the imperfect voltage sources
attached from outside to the quantum circuit.
Every Josephson junction (thick line) consists of $2$ branches:
a bare Josephson junction ($J$) and the junction capacitance ($C_J$)
as indicated in Fig.~\ref{fig:circuit}b.  
In addition to these two elements, a Josephson junction can also be
combined with a shunt resistance \cite{BKD}.  However, these resistances
are typically very large and can often be neglected;  they are not
be of interest here.
The circuit graph ${\cal G}$ is divided in two parts. 
The {\it tree} is a loop-free subgraph which connects all nodes of 
the circuit and it is represented by
solid lines in Fig.~\ref{fig:circuit}. 
All the branches $f_i$ ($i=1,\ldots,F$) that do not belong to the tree 
are called {\it chords} and are represented by dotted lines in
Fig.~\ref{fig:circuit}.
In the present case, the number of chords, not counting the
junction capacitances $C_J$, is $F=3$.
There can in principle be inductances contained both in the
tree and in the chords which considerably complicate the
analysis \cite{BKD}.  However, in our case there are no inductances 
in the tree (no $L$ inductances), so that our analysis is
much simpler than the general one.
 From now on, we make use of the fact that the 
circuit graph Fig.~\ref{fig:circuit} has no inductances in its tree.
When a chord is added to the tree, it gives 
rise to a unique loop, a  {\em fundamental loop}.
In other words, the set of fundamental loops 
${\cal  F}_i$ of the graph consists of all
loops which contain exactly one chord $f_i$.
The topological information about the graph is encoded in the  
fundamental loop matrix ${\bf F}^{(L)}$ of the circuit 
($i=1,\ldots,F$; $j=1,\ldots,B$),
\begin{equation}
  {\bf F}^{(L)}_{ij} = \left\{\begin{array}{l l}
      1, & \mbox{if $b_j \in {\cal F}_i$ (same direction)},\\
     -1, & \mbox{if $b_j \in {\cal F}_i$ (opposite direction)},\\
      0, & \mbox{if $b_j \notin {\cal F}_i$},
\end{array}\right. \label{circuit-rule}
\end{equation}
where the direction of the fundamental loop ${\cal F}_i$ is given
by the direction of its defining chord $f_i$. 
The currents ${\bf I}=(I_1,\ldots,I_B)$ and 
the voltages ${\bf V}=(V_1,\ldots,V_B)$ associated with the branches 
of the graph are
divided in into tree and chord currents and voltages,
\begin{equation}
{\bf I}=({\bf I}_{\rm{tr}},{\bf I}_{\rm{ch}}),\qquad
{\bf V}=({\bf V}_{\rm{tr}},{\bf V}_{\rm{ch}}).
\end{equation}
With the division into three and chord branches, the fundamental 
loop matrix assumes the block form
\begin{equation}\label{FundLoopMatrix}
{\bf F}^{(L)}=(-{\bf F}^T \, |\, \openone).
\end{equation}
We further split up the current and voltage vectors according
to the type of branch \cite{charge},
\begin{equation}
\begin{aligned}
{\bf I}_{\textrm{tr}}&=({\bf I}_J,{\bf I}_V,{\bf I}_Z), 
& {\bf I}_{\textrm{ch}}&=({\bf I}_{C_J},{\bf I}_C,{\bf I}_K),\\
{\bf V}_{\textrm{tr}}&=({\bf V}_J,{\bf V}_V,{\bf V}_Z), 
& {\bf V}_{\textrm{ch}}&=({\bf V}_{C_J},{\bf V}_C,{\bf V}_K),
\end{aligned}
\end{equation}
such that the matrix ${\bf F}$ acquires the sub-block form,
\begin{equation}
{\bf F}=\left(\begin{array}{ccc}
\openone & {\bf F}_{JC} & {\bf F}_{JK}\\
{\bf 0} & {\bf F}_{VC} & {\bf F}_{VK}\\
{\bf 0} & {\bf F}_{ZC} & {\bf F}_{ZK}\end{array}\right).
\end{equation}
By inspection of Fig.~\ref{fig:circuit}, one finds
the loop sub-matrices of the circuit according to the
rule Eq.~(\ref{circuit-rule}),
\begin{equation}
\begin{aligned}
{\bf F}_{JC} &= \left(\begin{array}{cc}  1 & 0\\ 0 & 1\\ 0 &
                     0\end{array}\right), &  
                 {\bf F}_{JK} &=
                     \left(\begin{array}{cc}  -1\\ 1\\
                     1\end{array}\right),\\ 
                 {\bf F}_{VC} = {\bf F}_{ZC}&=
                     \left(\begin{array}{cc}  1 & 0\\ 0 &
                     1\end{array}\right), &
                 {\bf F}_{VK}={\bf F}_{ZK} &=
                     \left(\begin{array}{cc}  0\\ 0\end{array}\right).
\end{aligned}
\label{loopmatrices}
\end{equation}
With Eq.~(\ref{FundLoopMatrix}), Kirchhoff's laws have the
compact form
\begin{eqnarray}
{\bf F}{\bf I}_{\textrm{ch}}&=&-{\bf I}_{\textrm{tr}},\\
{\bf F}^T{\bf V}_{\textrm{tr}}&=&{\bf V}_{\textrm{ch}}-\dot{\bf \Phi}_x,
\end{eqnarray}
where ${\bf \Phi}_x=(\Phi_1,\ldots,\Phi_F)$ is the vector of
externally applied fluxes.  Only loops with a non-zero inductance
are susceptible to an external magnetic flux, thus
only one external flux needs to be considered here,
${\bf \Phi}_x=(0,0,\Phi_x)$.

The SC phase differences across the junctions
$\bphi =(\varphi_1,\varphi_2,\varphi_3)$ are related
to the canonical variables, the fluxes ${\bf \Phi}$,
through the relation
\begin{equation}
\bphi=2\pi\frac{\bf \Phi}{\Phi_0},
\end{equation}
while the canonically conjugate momenta are the charges \
${\bf Q}=(Q_1, Q_2)$ on the junction capacitance.
Using circuit theory \cite{charge} and ignoring the
dissipative circuit elements $Z_1$ and $Z_2$ for the
moment, we find the following
Hamiltonian of the circuit Fig.~\ref{fig:circuit},
\begin{eqnarray}
{\cal H}_S &=&\frac{1}{2}\left({\bf Q}-{\bf C}_V {\bf V}_V\right)^T  {\cal
C}^{-1}\left({\bf Q}-{\bf C}_V {\bf V}_V\right)+{\cal U}({\bf \Phi}),\nn\\
\label{H}\\
{\cal U}({\bf \Phi})&=&
-{\bf E}_J {\bf cos}2\pi\frac{\bf \Phi}{\Phi_0}
+\frac{1}{2}\bf{\Phi}^T\bf{M}_0\bf{\Phi}+\bf{\Phi}^T\bf{N}\Phi_x,
\label{U} 
\end{eqnarray}
where we have defined 
${\bf cos}\bphi=(\cos\varphi_1,\cos\varphi_2,\cos\varphi_3)$.
The Josephson energy matrix is given as
\begin{equation}
{\bf E}_J = \left(\frac{\Phi_0}{2\pi}\right)^2 {\bf L}^{-1}_J
=  \textrm{diag}(E_J ,E_J ,\alpha E_J ),
\end{equation}
where $\Phi_0=h/2e$ is the SC quantum of magnetic flux.
We assume that the Josephson energies and capacitances 
of the junctions  $J_1$ and $J_2$ are equal, 
$E_{J1}=E_{J2}\equiv E_J$ and $C_{J1}=C_{J2}\equiv C_J$, 
and we define the ratio $\alpha=E_{J3}/E_J$. 
The capacitance matrices of the circuit are
\begin{equation}
{\bf C}_J = \textrm{diag}(C_J,C_J,C_{J3}),\quad  
{\bf C} = \textrm{diag}(C_1,C_2). \label{capacitances}
\end{equation}
The source voltage vector is defined as ${\bf V}_V=(V_1, V_2)$.  
The derived capacitance matrices ${\cal C}$ and ${\bf C}_V$ and the
derived (inverse) inductance matrices ${\bf M}_0$ and ${\bf N}$ of
Eq.~(\ref{H}) are given in the Appendix~\ref{sec:A}.



\section{Born-Oppenheimer approximation}
\label{sec:born-oppenheimer}

\begin{figure}[t]
   \begin{center}
     \includegraphics[width=8.0cm]{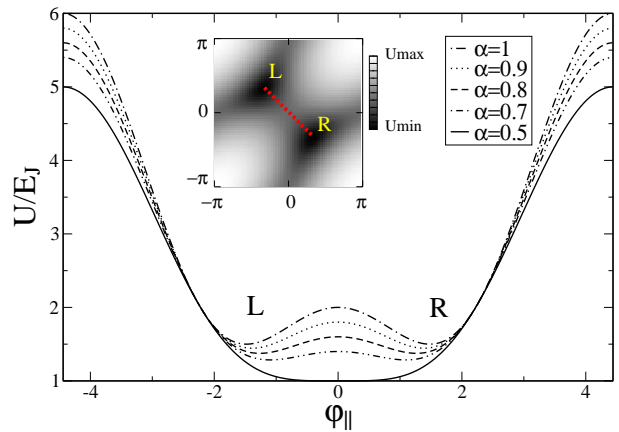}
     \caption{Plot of the potential ${\cal U}(\bphi)$ for $\varphi_x=\pi$
              along the line $\varphi_1+\varphi_2=0$ as a function of
              $\varphi_{\parallel}=\frac{1}{\sqrt{2}}
              (\varphi_1-\varphi_2)$ for several values of $\alpha$.
              In the curve for $\alpha=0.5$ the two minima are
              degenerate, while for $\alpha>0.5$ they split showing
              the double well.  The inset
              is a density plot of the potential for $\alpha=0.8$,
              showing the two minima and the line
              $\varphi_1+\varphi_2=0$.\label{fig:potential}}
   \end{center}
\end{figure}
We consider now
the limit in which the chord inductance $K$ is small  compared to the
Josephson inductances, $K\ll L_J$.
By means of the Born-Oppenheimer
approximation, we  derive an effective two-dimensional potential as a
function of two  ``slow'' degrees of freedom.
Our analysis follows closely that of \cite{DBK05}.  
For $K\ll L_J$, the potential Eq.~(\ref{U}) gives rise
to a hard constraint for the variables ${\bphi}$,
in the form of the linear equation
\begin{equation}\label{lineq}
{\bf M}_0\bphi+{\bf N}\varphi_x=0,
\end{equation}
where the external magnetic flux is written as  $\varphi_x =
2\pi\Phi_x/\Phi_0$.  
The general solution of the Eq.~(\ref{lineq}),
\begin{equation}
\bphi = \left(\begin{array}{c} \varphi_1\\ \varphi_2\\
\varphi_1-\varphi_2+\varphi_x
\end{array}\right),
\end{equation}
depends on the two variables $\varphi_1$ and
$\varphi_2$ only. Thus, in the limit of small $K$,  the
dynamics is restricted to a plane in three-dimensional  $\bphi$
space.   The potential,
restricted to the plane,  is then  a function of $\varphi_1$ and
$\varphi_2$ only \cite{Orlando99},
\begin{equation}
{\cal U}(\bphi)=E_J \Big[ -\cos(\varphi_1)-
\cos(\varphi_2)-\alpha\cos(\varphi_1-\varphi_2+\varphi_x)
\Big].\label{U2}
\end{equation}
A density plot of ${\cal U}$ for $\alpha=0.8$ as a function 
of $\varphi_1$ and $\varphi_2$ is shown in the inset of 
Fig.~\ref{fig:potential}.
The minima of the potential are found by solving the
equation ${\grad}\,{\cal U}=0$,
which yields \cite{Orlando99}
\begin{equation}
\sin\varphi_1=-\sin\varphi_2=-\sin\varphi^*,
\end{equation}
where $\varphi^*$ is the solution of the self-consistent equation
\begin{equation}\label{eq:self-consistent}
\sin\varphi^*=\alpha\sin(2\varphi^*+\varphi_x).
\end{equation}
The potential forms two wells whose relative depth is determined  by the
value of the externally applied flux $\varphi_x$.  In order to have a
symmetric double well we choose $\varphi_x=\pi$ which yields two minima
at the points $\bphi_{\rm R}=(\varphi^*,-\varphi^*)$  and
$\bphi_{\rm L}=(-\varphi^*,\varphi^*)$
with  $\varphi^*=\arccos(1/2\alpha)>0$. If $\alpha>0.5$, then there 
are two distinct minima. Taking into account the periodicity of the
potential, a complete set of solutions of Eq.~(\ref{eq:self-consistent}) is
$\bphi=\pm(\varphi^*,-\varphi^*)^T+2\pi(n,m)$, with integer $n,m$.  
We plot the double well
potential between the two minima in Fig.~\ref{fig:potential} for
different values of $\alpha$  in the symmetric case $\varphi_x=\pi$.


\section{Quantum dynamics}
\label{sec:quantum-dynamics}

In this section, we look for localized solutions of the Schr\"odinger
equation ${\cal H} \Psi = E \Psi$, with the Hamiltonian of
Eq.~(\ref{H}). 
We expand the potential around the two
minimum  configurations, keeping contributions up to the second order
in $\bphi$,  and solve the Schr\"odinger equation in these two
different points (denoting them L and R for left and right).  
We obtain the quadratic Hamiltonian
\begin{equation}\label{QuadHam}
{\cal H}_{\rm L,R}=\frac{1}{2}\left[{\bf Q}^T   {\cal C}^{-1}{\bf Q}+
\boldsymbol{\Phi}^T
{\bf L}_{\textrm{lin; L,R}}^{-1}\boldsymbol{\Phi}\right],
\end{equation}
where the linearized inductance ${\bf L}_{{\rm lin; L,R}}$ is
defined as
\begin{equation}
{\bf L}_{\textrm{lin; L,R}}^{-1}
={\bf M}_0
+{\bf L}_J^{-1}{\bf cos}\bphi_{\rm L, R}.
\end{equation}
To simplify the kinetic part in Eq.~(\ref{QuadHam}),
we perform a canonical transformation on the variable  
$\boldsymbol{\Phi}$ and its conjugate momentum ${\bf Q}$ \cite{DBK05},
\begin{eqnarray}
\boldsymbol{\Phi} &=& \sqrt{c} \left(\sqrt{{\cal C}}^{-1}\right)^T 
\tilde{\boldsymbol{\Phi}},\nn\\ 
{\bf Q} &=& \sqrt{{\cal C}}\tilde{{\bf Q}}/\sqrt{c},
\end{eqnarray}
where $c$ is an arbitrary unit capacitance (e.g., $c=C_J$).
We define the diagonal matrix ${\bf \Omega}^2_{L,R}$ such that
it satisfies
\begin{equation}\label{Omega2}
(\sqrt{{\cal C}}^{-1})^T {\bf L}_{\textrm{lin;L,R}}^{-1}  
\sqrt{{\cal C}}^{-1}={\bf O}^T{\bf \Omega}^2_{L,R}{\bf O},
\end{equation}
where ${\bf O}$ is an orthogonal matrix that diagonalizes the left
hand side (lhs) of Eq.~(\ref{Omega2}).  
This allows us to further simplify the Hamiltonian by
making the following  canonical transformation, preserving the Poisson
brackets,
\begin{equation}
\boldsymbol{\Phi}'  = {\bf O} \tilde{\boldsymbol{\Phi}}, \quad
{\bf Q}'  = {\bf O}\tilde{{\bf Q}},
\end{equation}
that leads us to the Hamiltonian,
\begin{equation}
{\cal H}_{\rm L,R}=\frac{1}{2}\left(c^{-1}{\bf Q}'^2+\boldsymbol{\Phi'}^T
{\bf \Omega}^2_{\rm L,R}\boldsymbol{\Phi'}\right).
\end{equation}
In the case of a symmetric potential (when $\varphi_x=\pi$),  the
matrices $\bf{L}_{\textrm{lin;L,R}}$  of the linearized problem
are equal,
\begin{equation}
{\bf L}_{\textrm{lin;L}} = {\bf L}_{\textrm{lin;R}} ,
\quad{\rm and}\quad {\bf \Omega}_{\rm L} = {\bf \Omega}_{\rm R} ,
\end{equation}
hence we drop the subscript L and R for simplicity.

\begin{figure}[t]
   \begin{center}
     \includegraphics[width=8.0cm]{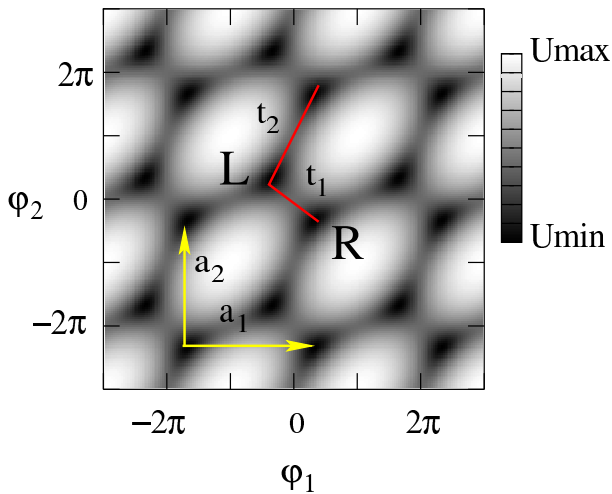}
     \caption{Density plot of the double well potential
              $U(\varphi_1,\varphi_2)$ in units of $E_J$ for
              $\alpha=0.8$ on a logarithmic scale.  The
              periodicity of the potential is evident;  each unit cell
              contains two minima (black).  
	      The primitive vectors of the Bravais lattice are 
              denoted ${\bf a}_1$ and
              ${\bf a}_2$ while $t_1$  and $t_2$
              are the tunneling matrix elements between the
              nearest-neighbor minima. \label{fig:lattice}}
   \end{center}
\end{figure}

We quantize the Hamiltonian by imposing the canonical
commutation  relations,
\begin{equation}
\left[\Phi_i,Q_j\right]=i\hbar\delta_{ij},
\end{equation}
where $\Phi_i$ and $Q_j$ are the components of the vectors  ${\bf
\Phi}$ and ${\bf Q}$ respectively.
The ground-state wave function is the Gaussian,
\begin{equation}
\Psi_{\alpha}(\boldsymbol{\varphi})=
\left(\frac{\det{\cal M}}{\pi^2}\right)^{1/4}
\exp\left[-\frac{1}{2}
(\boldsymbol{\varphi}-\boldsymbol{\varphi}_{\alpha})^T {\cal M}
(\boldsymbol{\varphi}-\boldsymbol{\varphi}_{\alpha})\right],
\label{Psi}
\end{equation}
where $\alpha={\rm L,R}$ and 
\begin{equation}\label{M}
{\cal M}=\frac{1}{\hbar}\left(\frac{\Phi_0}{2\pi}\right)^2\sqrt{\cal C}{\bf
  O}^T{\bf \Omega}{\bf O}\sqrt{\cal C}.
\end{equation}
For the wave function overlap integral between the  left and
right state, $S=\langle \Psi_L|\Psi_R\rangle$, we find
\begin{equation}
S=\exp\left\{-\frac{1}{4}
\boldsymbol{\Delta\varphi}^T{\cal M}\boldsymbol{\Delta\varphi}\right\},
\end{equation}
where $\Delta \bphi =\bphi_{\rm R} -\bphi_{\rm L}=2\arccos(1/2\alpha)(1,-1)$ 
is the distance between
the right (R) and left (L) potential minima (Fig.~\ref{fig:lattice}).


\section{Bloch theory}
\label{sec:bloch-theory}

Given the periodicity of the problem Eq.~(\ref{H})
with the potential Eq.~(\ref{U2}) in the Born-Oppenheimer
approximation, an important question concerns the boundary 
conditions of the problem, i.e., the choice of the
appropriate Hilbert space.  The question is whether
the domain of $\bphi$ should be the infinite plane
or the square $T=[-\pi,\pi)^2$ with periodic boundary
conditions.
This question has been discussed extensively in the literature
\cite{Averin85,Likharev85,Iansiti89}.
Since in our case, 
a shift of $\varphi_1$ or $\varphi_2$ by $2\pi$ creates a state 
which is physically indistinguishable from the one before the shift,
we choose the compact domain $T$ and impose periodic boundary
conditions on the wavefunction. 
However, we are going to extend the domain to the
infinite domain in order to facilitate the calculation.

\subsection{The periodic problem}
\label{ssec:periodic-problem}

The approximate solutions constructed in Sec.~\ref{sec:quantum-dynamics}
are a good starting point, but they are insensitive to the boundary
conditions.  
However, the boundary conditions are essential if finite bias
voltages ${\bf V}_V$ are to be taken into account.
The problem at hand is defined on the square with side $2\pi$
(see inset of Fig.~\ref{fig:potential}) with periodic boundary
conditions;
i.e., the phases $\boldsymbol {\varphi}=(\varphi_1,\varphi_2)^T$ 
are in the compact domain $T=[-\pi,\pi)^2$ and the wavefunction
at opposite edges needs to be identical,
$\Psi(-\pi,\varphi_2)=\Psi(\pi,\varphi_2)$
and $\Psi(\varphi_1,-\pi)=\Psi(\varphi_1,\pi)$,
such that $T$ acquires the topology of a torus.
If the boundary conditions are ignored, e.g., in the case
where the wavefunction is known to be vanishingly small at the 
boundary, then the bias voltages ${\bf V}_V$ in 
the Hamiltonian Eq.~(\ref{H}) can be removed completely with a gauge 
transformation and the solutions will be independent of ${\bf V}_V$.

We proceed as follows:  We first solve the problem Eq.~(\ref{H}) 
in the infinite two-dimensional plane and then choose those solutions
that satisfy the periodic boundary conditions and then restrict
them to the compact domain $T$.  We choose this
approach because the problem on the infinite domain is
well known:  the solutions 
$\psi_{\alpha{\bf k}}$ are given by Bloch's theorem
for the motion of a particle in a crystal and satisfy
\begin{equation}
\psi_{\alpha{\bf k}}(\boldsymbol{\varphi}+2\pi{\bf m}) =\e^{2\pi i{\bf
m}\cdot{\bf k}} \psi_{\alpha{\bf k}}(\boldsymbol{\varphi}),
\end{equation}
for ${\bf m}=(m_1,m_2)$ with integer $m_1$ and $m_2$.
The minima of our potential, Eq.~(\ref{U2}), define a 
two-dimensional square Bravais lattice 
with a two-point basis, which looks like a sheared hexagonal
lattice (although it is a square lattice).
The lattice and its primitive vectors 
${\bf a}_1=(2\pi,0)$ and ${\bf a}_2=(0,2\pi)$ are shown in
Fig.~\ref{fig:lattice}.
The lattice basis is given by the vectors
${\bf b}_{\rm L}=(0,0)$ and  ${\bf b}_{\rm R}=2(\varphi^*,-\varphi^*)$.
Each lattice point can be identified by the Bravais lattice
vector ${\bf n}$ and the basis index $\alpha={\rm L,R}$.
As indicated above, not all the Bloch functions satisfying the 
Schr\"odinger equation on the infinite domain have a physical meaning,  
but only those that are also $2\pi$-periodic.  In the case of zero applied
voltage bias, the only value of ${\bf k}$  yielding to a periodic wave
function is ${\bf k}={\bf 0}$.

\subsection{Tight-binding approximation}
\label{ssec:tight-binding}

In order to construct approximate Bloch states, we first form
localized Wannier orbitals $\phi_\alpha$ by orthonormalizing the
localized solutions $\Psi_\alpha$ ($\alpha={\rm L,R}$) from
Eq.~(\ref{Psi}).
These Wannier orbitals are centered at arbitrary lattice points,
$\phi_{\alpha{\bf n}}(\bphi)=\phi_\alpha(\bphi-2\pi{\bf n})$
and satisfy the orthonormality relations
\begin{equation}
\langle\phi_{\alpha{\bf n}}|\phi_{\beta{\bf m}}\rangle
=\delta_{\alpha \beta}\delta_{\bf n m}.
\end{equation}
The Bloch states are then related to the Wannier orbitals via a
Fourier transform,
\begin{eqnarray}
\psi_{\alpha{\bf k}}(\boldsymbol{\varphi}) &=&
\sum_{{\bf n}\in{\bb Z}^2} \e^{2\pi i{\bf k}\cdot{\bf n}}
\phi_{\alpha{\bf n}}
(\boldsymbol{\varphi}),\label{FT}\\
\phi_{\alpha{\bf n}}(\boldsymbol{\varphi}) &=&
\int_{\rm FBZ}d{\bf k} \e^{-2\pi i{\bf k}\cdot{\bf n}} 
\psi_{\alpha{\bf k}}(\boldsymbol{\varphi}),\label{IFT}
\end{eqnarray}
where the integration in Eq.~(\ref{IFT}) is over the first Brillouin 
zone (FBZ), i.e., $k_i\in[-1/2,1/2)$.
The label $\alpha$ plays the role of the energy band label in Bloch 
theory.
The Bloch states $\psi_{\alpha{\bf k}}$ form a complete set  of
orthonormal states in ${\bf k}$-space, where $k_i\in[-1/2,1/2)$,
\begin{eqnarray}
&&\langle\psi_{\alpha{\bf k}}| \psi_{\beta{\bf q}}\rangle
=\delta_{\alpha\beta} \delta({\bf k}-{\bf q}),\label{Bloch-orth}\\
&&\sum_{\alpha}\int d{\bf k}|\psi_{\alpha{\bf k}}\rangle
\langle\psi_{\alpha{\bf k}}|=\openone.
\label{Bloch-compl}
\end{eqnarray}
For the completeness relation Eq.~(\ref{Bloch-compl}) to hold, we must
sum over all bands $\alpha$, corresponding to a complete set of Wannier
functions.  Here, in order to describe the low-energy physics of the
system, we restrict ourselves to the two lowest bands $\alpha={\rm L,R}$,
related to the left and right potential minimum in the unit cell,
and neglect higher excited states of the double wells.  
This restriction
is justified if the energy gap between the lowest two states is much 
smaller than the gap between the two lowest and all higher states
(see Table \ref{tab}).
We normalize the Bloch functions on the unit cell $T$,
\begin{equation}
\int_T d\bphi|\psi_{{\bf k}\alpha}(\boldsymbol{\varphi})|^2=1.
\end{equation}

Now we can expand the Hamiltonian in the Bloch function basis with
Eq.~(\ref{Bloch-compl}), and then apply Eq.~(\ref{FT}),
\begin{eqnarray}
{\cal H} & \simeq & \sum_{\alpha\beta}\int d{\bf k} d{\bf q} |\psi_{\alpha
{\bf k}}\rangle\langle\psi_{\alpha{\bf k}} |{\cal H}| 
\psi_{\beta{\bf q}}\rangle\langle\psi_{\beta{\bf q}}|\nn\\ 
&=&\sum_{\alpha\beta}\int d{\bf k}d{\bf q}
\,{\cal H}^{\alpha\beta}_{\bf k q}\,
|\psi_{\alpha{\bf k}}\rangle\langle\psi_{\beta{\bf q}}|,
\end{eqnarray}
where the approximation in the first line consists in omitting
bands that are energetically higher than $\alpha={\rm L,R}$ (see above).
The matrix elements of the Hamiltonian in the Bloch basis are
\begin{equation}
{\cal H}^{\alpha\beta}_{\bf k q}= \sum_{{\bf n,m}\in{\bb Z}^2}
\e^{-2\pi i({\bf k}\cdot{\bf n}-{\bf q}\cdot{\bf m})}
\langle\phi_{\alpha{\bf n}}|{\cal H} |\phi_{\beta{\bf m}}\rangle.
\label{Hamiltonian-kq}
\end{equation}
For fixed ${\bf k}$ and ${\bf q}$, 
Eq.~(\ref{Hamiltonian-kq}) is reduced to a $2\times 2$ hermitian
matrix. 
The main contributions to Eq.~(\ref{Hamiltonian-kq}) stem from 
either tunneling between the two sites in the same unit cell
(intra-cell) or between site L in one cell and site R in an adjacent
cell (inter-cell), see Fig.~\ref{fig:lattice}.
For the off-diagonal element we can write
\begin{eqnarray}
{\cal H}^{\rm LR}_{\bf kq}&\simeq& \sum_{{\bf n}\in{\bb Z}^2} 
e^{-2\pi i({\bf k}-{\bf q})\cdot {\bf n}}
\Big[\langle\phi_{{\rm L}{\bf n}}|{\cal H} |\phi_{{\rm R}{\bf n}}\rangle  \nn\\
& &
 + \e^{-2\pi i q_1} \langle\phi_{{\rm L}{\bf n}}|{\cal H} 
|\phi_{{\rm R}{\bf n-e_1}}\rangle \nn\\ 
& & + \e^{2\pi i q_2}\langle\phi_{{\rm L}{\bf n}}|{\cal H} 
|\phi_{{\rm R}{\bf n+e_2}}\rangle\Big].\label{HLR}
\end{eqnarray}
where ${\bf e}_1=(1,0)$ and  ${\bf e}_2=(0,1)$.
Due to the lattice periodicity, 
the quantities (see Fig.~\ref{fig:lattice})
\begin{eqnarray}
\epsilon_0&=&\langle\phi_{L(R){\bf n}}|{\cal H}|\phi_{L(R){\bf
n}}\rangle,\label{eps0}\\ 
t_1&=&\langle\phi_{L(R){\bf n}}|{\cal H}|\phi_{R(L){\bf n}}\rangle,\label{t1}\\ 
t_2&=&\langle\phi_{L(R){\bf n}}|{\cal H}|\phi_{R(L){\bf n-e_1}}\rangle\\ 
&=& \langle\phi_{L(R){\bf n}}|{\cal H}|\phi_{R(L){\bf n+e_2}}\rangle,\label{t2}
\label{NeighTunn}
\end{eqnarray}
are independent of the lattice site ${\bf n}$, and thus from
Eq.~(\ref{Hamiltonian-kq}), we find ${\cal H}_{\bf
  kq}^{\alpha\beta}\simeq\delta({\bf k}-{\bf q}){\cal H}_{\bf
  k}^{\alpha\beta}$. 
We can now write the $2\times 2$ Hamiltonian as
\begin{eqnarray}
{\cal H}_{\bf k}&=& \epsilon_0 \openone+
\frac{1}{2}\left(\begin{array}{cc}0 & \Delta({\bf
k})^*\\ \Delta({\bf k}) & 0\label{H2}\\
\end{array}\right),\\
\Delta({\bf k})&=& 2\left[t_1+t_2(\e^{2\pi i k_1} +\e^{-2\pi i  k_2})\right]\label{delta}.
\end{eqnarray}
The equality in Eq.~(\ref{NeighTunn}) is due to the invariance of the
potential under the transformation
$(\varphi_1,\varphi_2) \rightarrow -(\varphi_2,\varphi_1)$ 
and it is valid also in the $\varphi_x\ne\pi$ case.
The eigenvalues of the problem are
\begin{equation}
\epsilon_{\pm}({\bf k})
=\epsilon_0\pm \frac{1}{2}|\Delta({\bf k})|,
\label{eigenenergy}
\end{equation}
and represent a typical two-band
dispersion relation. 
In the case of zero external applied voltage
only the  ${\bf k}=0$ Bloch functions satisfy the correct boundary
conditions, i.e., are periodic. 
For ${\bf k}=0$ we recognize the
qubit Hamiltonian that, in the symmetric  double well case, is given
by a $\sigma_x$ term \cite{Orlando99},
\begin{equation}
{\cal H}=\epsilon_0+(t_1+2t_2)\sigma_x.
\end{equation}

\subsection{Effect of a Voltage bias}
\label{ssec:voltage-bias}

Now, we study the case with an (nonzero) external bias voltage.
Given the Bloch function $\psi_{\alpha{\bf k}}$ that satisfies 
the Schr\"odinger equation
for the Hamiltonian Eq.~(\ref{H}) for zero applied voltages,  ${\bf V}_V=0$,
we find for the solution wave function for finite voltages ${\bf V}_V \ne 0$,
\begin{equation}
u_{\alpha{\bf k}}(\boldsymbol{\varphi})= 
\e^{-i\boldsymbol{\varphi}\cdot{\bf Q}_g/2e} 
\psi_{\alpha{\bf k}}(\boldsymbol{\varphi}),
\label{wavefunction-shift}
\end{equation} 
where we have defined the gate charge vector as ${\bf Q}_g={\bf C}_V {\bf V}_V$.
The above statement can be directly verified by substituting 
$u_{\alpha{\bf k}}$
from Eq.~(\ref{wavefunction-shift}) into the Schr\"odinger equation
with Eq.~(\ref{H}) while using that $\psi_{\alpha{\bf k}}$ solves the problem for
${\bf V}_V=0$.
The solutions in the presence of an applied voltage bias satisfy
\begin{equation}
u_{\alpha{\bf k}}(\boldsymbol{\varphi}+2\pi{\bf n})= \e^{2\pi i{\bf
n}\cdot({\bf k}-{\bf Q}_g/2e)} u_{{\alpha}\bf k}(\boldsymbol{\varphi}).
\end{equation}
For the periodicity of the wave function on the compact domain, we
have to choose ${\bf k}={\bf Q}_g/2e$.
This means that $u_{\alpha{\bf k}}$ 
is the periodic part of the Bloch function
for ${\bf k}={\bf Q}_g/2e$.
By substituting this into Eqs.~(\ref{H2}) and (\ref{delta}),
we obtain the qubit Hamiltonian
\begin{equation}
{\cal H} = \frac{1}{2}\left[\textrm{Re}(\Delta) \sigma_x
  + \textrm{Im}(\Delta)\sigma_y
  + \epsilon \sigma_z\right]
  = \frac{1}{2}{\bf B}\cdot {\bsigma},
\label{qubitH}
\end{equation}
where we have also included the effect of a (small) bias flux
that tilts the double well, 
$\epsilon\simeq 2\alpha \sqrt{1-1/4\alpha^2} E_J (\varphi_x-\pi)$,
where $\bsigma=(\sigma_x, \sigma_y, \sigma_z)$ are the 
Pauli matrices, and
\begin{eqnarray}
\textrm{Re}(\Delta) & = &  2\left[t_1+2t_2\cos(\pi k_+)\cos(\pi k_-)\right],\label{ReD}\\ 
\textrm{Im}(\Delta) & = &  4t_2\cos(\pi k_+)\sin(\pi k_-),\label{ImD}
\end{eqnarray}
with $k_\pm = (C_1 V_1 \pm C_2 V_2)/2e$.
The eigenstates for $\epsilon=0$ are
\begin{eqnarray}
|0\rangle&=&\frac{1}{\sqrt{2}} \left(-\e^{-i\theta}
|L\rangle +|R\rangle\right),\\
|1\rangle&=&\frac{1}{\sqrt{2}} \left(\e^{-i\theta}
|L\rangle +|R\rangle \right),
\end{eqnarray}
where $\tan\theta=\textrm{Im}(\Delta)/\textrm{Re}(\Delta)$.  
In Eq.~(\ref{qubitH}), we have introduced the pseudo-field
${\bf B}=(\textrm{Re}(\Delta),\textrm{Im}(\Delta),\epsilon)$.

\section{Calculation of $t_1$ and $t_2$}
\label{sec:t1t2}

For a quantitative analysis of the single-qubit Hamiltonian Eq.~(\ref{qubitH}),
we have to calculate the tunneling matrix elements $t_1$ and $t_2$.
In order to do so, we require a set of orthonormal Wannier functions on
the infinite two-dimensional lattice defined by the potential ${\cal U}$,
Eq.~(\ref{U}).
We start from the non-orthogonal set of Gaussian orbitals
$|\Psi_{\alpha{\bf n}}\rangle$ consisting of the solution
Eq.~(\ref{Psi}), shifted by a lattice vector ${\bf n}$,
\begin{equation}
\Psi_{\alpha{\bf n}}(\bphi) = \Psi_{\alpha}(\bphi-2\pi{\bf n}).
\end{equation}
The orthonormalized Wannier functions can be written 
as a linear combination of these Gaussians,
\begin{equation}
|\phi_{\alpha{\bf n}}\rangle=\sum_{\mu=L,R,{\bf l}\in{\bb Z}^2}
{\cal G}_{\mu{\bf l},\alpha{\bf n}}|\Psi_{\mu{\bf l}}\rangle.
\end{equation}
To form a complete set of orthonormal functions the
following relation must be satisfied,
\begin{equation}
\langle\phi_{\alpha{\bf n}}|\phi_{\beta{\bf m}}\rangle
 = \left({\cal G}^{\dag}S{\cal G}\right)_{\alpha{\bf n},\beta{\bf m}}
=\delta_{\alpha \beta} \delta_{{\bf n}{\bf m}},
\label{G-matrix}
\end{equation}
where $S$ is the (real and symmetric) overlap matrix,
\begin{equation}
\label{S-matrix}
S_{\alpha{\bf n},\beta{\bf m}}=\int d\boldsymbol{\varphi}
\Psi_{\alpha{\bf n}}(\boldsymbol{\varphi})
\Psi_{\beta{\bf m}}(\boldsymbol{\varphi}).
\end{equation}
We solve Eq.~(\ref{G-matrix}) with
\begin{equation}
\label{GinverseS}
{\cal G}^T={\cal G}=\sqrt{S^{-1}}.
\end{equation}
The inverse of $S$ exists due to its positive definiteness.
The entries of the overlap matrix $S$ are equal to 1 on the diagonal, whereas
the off-diagonal elements are positive and $\ll 1$ because the
orbitals $\Psi_{\alpha{\bf n}}$ are well localized.
We define the matrix $S^{(1)}$ with all matrix elements $\ll 1$ via
\begin{equation}
S = \openone+S^{(1)} = \openone
+\left(\begin{array}{cc} S_{\rm LL} & S_{\rm LR} \\ S_{\rm LR}^T & S_{\rm RR}
\end{array}\right),
\end{equation}
and find, keeping only first order terms in $S^{(1)}$,
\begin{equation}
{\cal G}\simeq\sqrt{S^{-1}}\simeq \openone-\frac{1}{2}S^{(1)}.
\end{equation}
Note that $S_{\rm LL}$ and $S_{\rm RR}$ have zeros on the diagonal.

\begin{figure}[t]
   \begin{center}
     \includegraphics[width=8cm]{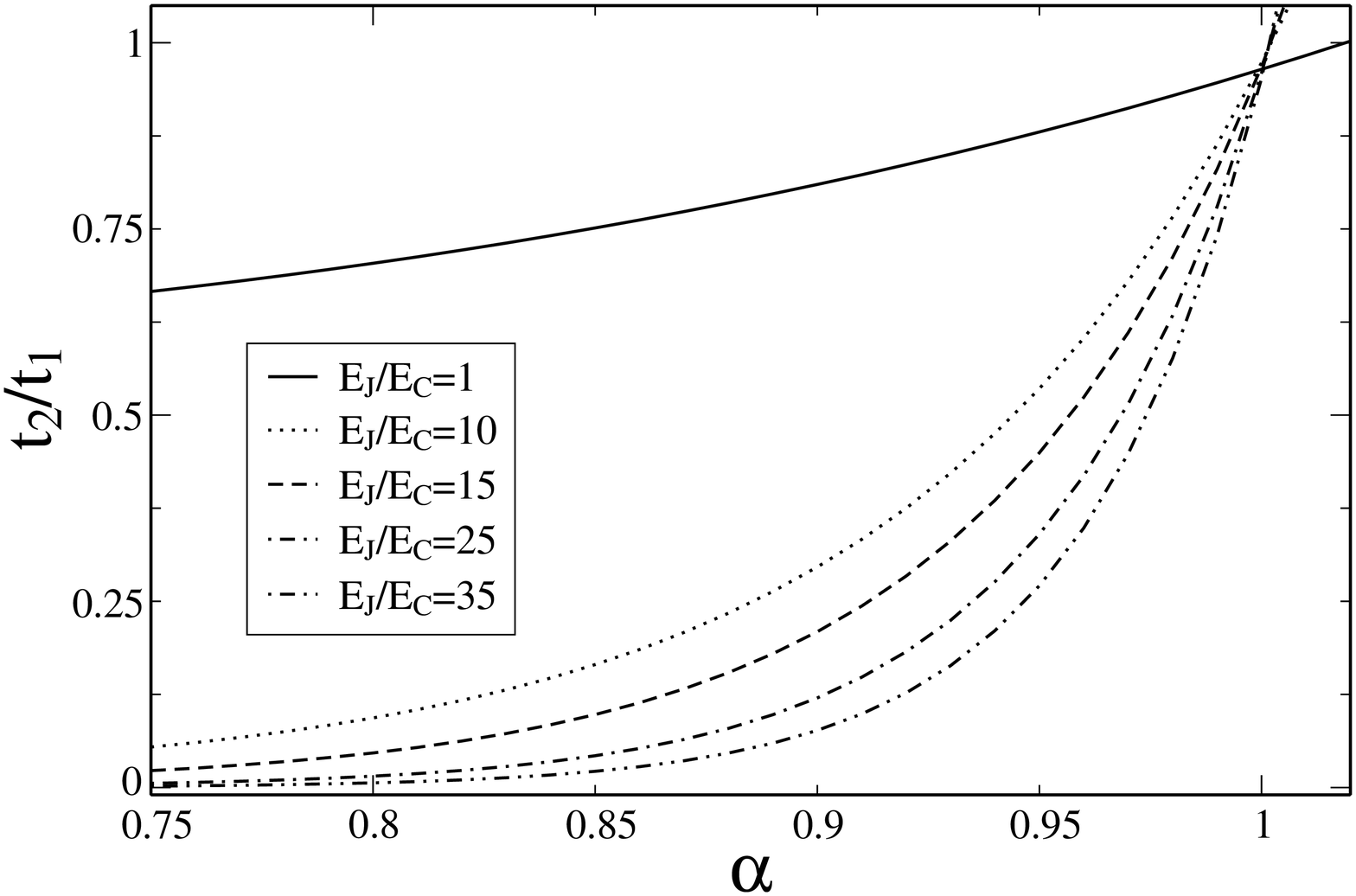}
     \caption{\small The ratio $t_2/t_1$ between the tunneling matrix
       elements, plotted as a function of
             $\alpha\le 1$ for  several values of
             $E_J/E_C$.\label{fig:t21vsAlpha}}
   \end{center}
\end{figure}
In our tight-binding approximation, we consider five unit cells, 
a center cell with its four nearest neighbors, corresponding
to the lattice vectors $\{(0,0), (\pm 1,0), (0,\pm 1)\}$.
This means that $S$ and ${\cal G}$ are $10\times 10$ matrices,
which can also be expressed as $2\times 2$ block matrices, each block 
of dimension $5\times 5$.
The two largest values are given by $s_1=S_{{\rm L}{\bf n},{\rm R}{\bf n}}$
and $s_2=S_{{\rm L}{\bf n},{\rm R}{\bf n}-{\bf e}_1}
=S_{{\rm L}{\bf n},{\rm R}{\bf n}+{\bf e}_2}$ \
with the nearest neighbor cell.  
Taking only these two largest overlaps into account,
we obtain $S_{\rm LL}=S_{\rm RR}\simeq 0$ and
\begin{equation}
S_{\rm LR}\simeq\left(\begin{array}{ccccc} s_1 & s_2 & s_2 & 0 & 0\\ 0 & s_1 &
0 & 0 & 0\\ 0 & 0 & s_1 & 0 & 0\\ s_2 & 0 & 0 & s_1 & 0\\ s_2 & 0 & 0
& 0 & s_1\end{array}\right).
\end{equation}

\begin{figure}[t]
   \begin{center} 
     \includegraphics[width=7.7cm]{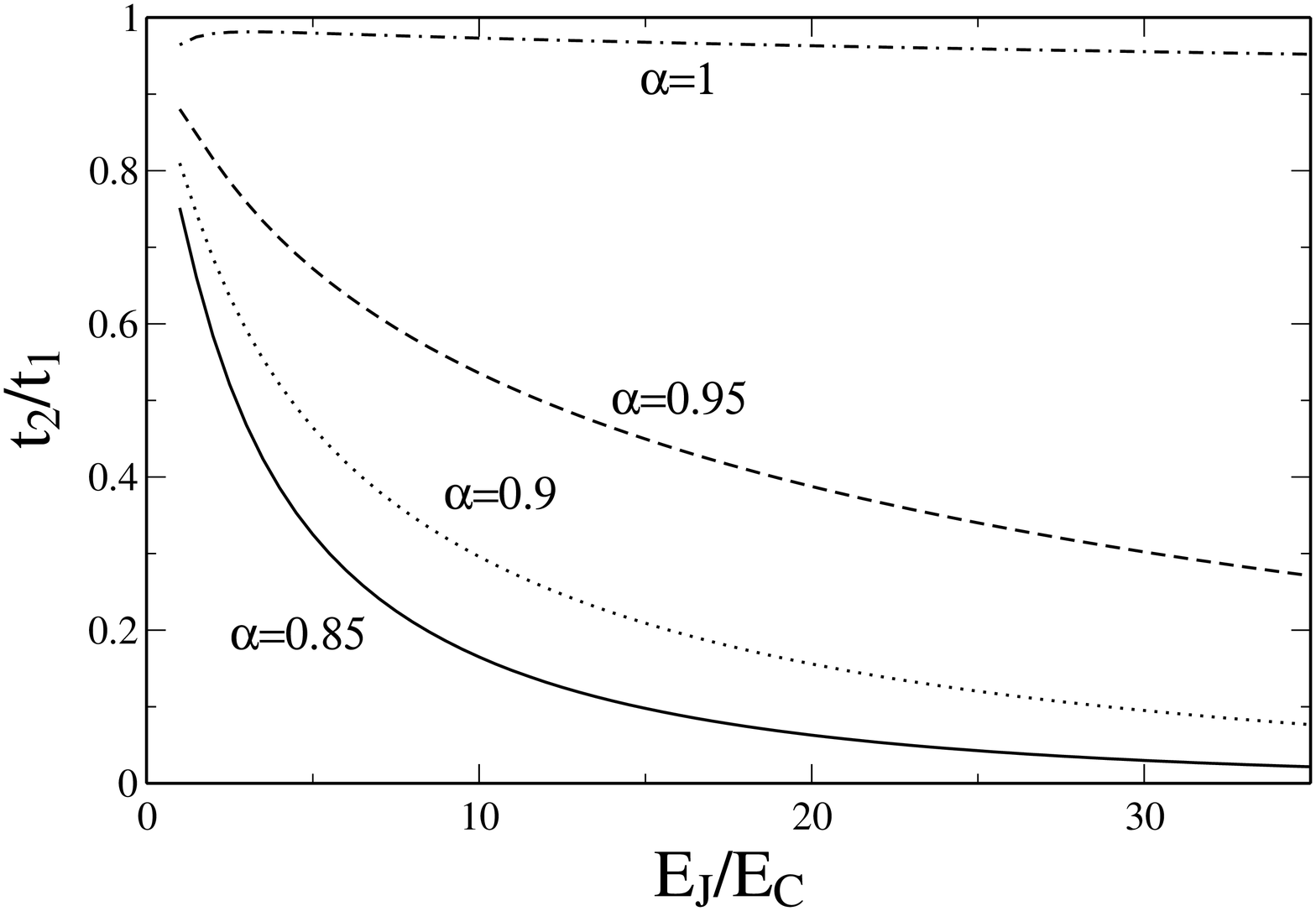}
     \caption{\small The ratio $t_2/t_1$ between the tunneling matrix elements,
       plotted as a function of
             $E_J/E_C$  for several values of
             $\alpha\le 1$.\label{fig:t21vsTau}}
   \end{center}
\end{figure}
Having the matrix ${\cal G}$ and $S$ we can calculate the tunneling
matrix
\begin{equation}
{\cal T}_{\alpha{\bf n},\beta{\bf m}}=\langle\phi_{\alpha{\bf
    n}}|{\cal H} |\phi_{\beta{\bf m}}\rangle
=({\cal G}^{\dag}T{\cal G})_{\alpha{\bf
    n},\beta{\bf m}},\label{Tcal}
\end{equation}
where the entries of the matrix $T$ are given as
\begin{equation}
T_{\alpha{\bf n},\beta{\bf m}}
= \langle \Psi_{\alpha{\bf n}}|{\cal H}|\Psi_{\beta{\bf m}}\rangle.
\label{T}
\end{equation}
Since both the $|\Psi_{\alpha{\bf n}}\rangle$ and the
$|\phi_{\alpha{\bf n}}\rangle$ states are localized at
the lattice position ${\bf n}$, the matrices $T$ and
${\cal T}$ both have the same non-zero entries as $S$.
The tunneling matrix ${\cal T}$ has the same block form
as $S$ with
${\cal T}_{\rm LL}={\cal T}_{\rm RR}=\epsilon_0\openone$ and
${\cal T}_{\rm LR}$ having the same structure
as $S_{\rm LR}$ with $s_1$ and $s_2$ replaced by
$t_1$ and $t_2$, given as $t_1={\cal T}_{{\rm L}{\bf
n},{\rm R}{\bf n}}$  and $t_2={\cal T}_{{\rm L}{\bf n},{\rm R}{\bf n}-{\bf e}_1}
={\cal T}_{{\rm L}{\bf n},{\rm R}{\bf n}+{\bf e}_2}$. 
The overlaps $s_1$ and $s_2$, together with the transition amplitudes
$t_1$ and $t_2$, depend exponentially on the two parameters $\alpha$ and
$E_J/E_C$. 
A detailed analysis is given below;
here, we anticipate the approximate relations
$t_1/t_2 >1$ if $\alpha<1$,
$t_1/t_2 <1$ if $\alpha>1$,
and $t_1/t_2 \approx 1$ if $\alpha=1$, and $t_1/t_2=1$ if $C_1=C_2=0$.

Now, we numerically determine the tunneling matrix elements
$t_1$ and $t_2$ from Eqs.~(\ref{Tcal}) and (\ref{T})
and analyze their dependence on the external parameters.
This dependence can then be used to control the qubit
Hamiltonian.
The external parameters fall into two categories, those that
can be varied freely, like magnetic fields and bias voltages, and the
device parameters, that are fixed for a specific device. 
Two main types of device parameters characterize the Hamiltonian:
(i) the junction capacitance $C_J$ that determines the charging energy 
$E_C=e^2/2C_J$ and (ii) the Josephson inductance $L_J$ which determines
the Josephson energy  $E_J=(\Phi_0/2\pi)^2/L_J$. In addition, we have
the ratio $\alpha=E_{J3}/E_J$.

The potential ${\cal U}(\bphi)$ can be modified in two ways. 
The external magnetic flux $\Phi_x=\Phi_0\varphi_x/2\pi$
is responsible for the symmetry  of the double well within a unit cell
and can give rise to a $\sigma_z$ term  in the single qubit Hamiltonian
while 
$\alpha$ determines the height of the barrier  between
the wells in a cell and between two nearest neighbor unit cells. Thus
$\alpha$ affects the values of the tunneling amplitudes between different
sites in the lattice. Although $\alpha$ is a fixed device parameter
for the set-up shown in Fig.~\ref{fig:schematic},  a modified set-up
in which the middle junction is made flux-tunable has been proposed
\cite{Orlando99,Makhlin01}; a flux tunable junction is achieved by
``shunting'' the third junction with a further junction and using an
external magnetic field to tune it. 

In the tight-binding picture, the off-diagonal element $\Delta$ of 
the qubit Hamiltonian is a complex quantity that depends  on
the two tunneling amplitudes $t_1$ and $t_2$, whose relative strength
can be set by $\alpha$ and the ratio $E_J/E_C$. The latter enters as
a common factor into the frequencies of the Gaussian localized
orbitals, determining the size of their overlaps and affecting only
the energy gap $|\Delta|$.
An increase of the value of $\alpha$ implies a
decrease of the  tunneling amplitudes $t_1$ and $t_2$, caused by an
increase of the height  of the barriers. Thus a careful choice of the
two parameters is crucial in determining the behavior of the
system.
 From Eq.~(\ref{delta}), we find that if $t_2/t_1\ll 1$ then $\Delta$ will be
(almost) real. In order to obtain a sizable imaginary part of $\Delta$,
$t_2/t_1$ must be sufficiently large.
In Fig.~\ref{fig:t21vsAlpha}, we plot the ratio $t_2/t_1$ versus
$\alpha$, for several values of the $E_J/E_C$.
Although all the curves
approach the value $t_2/t_1\approx 1$ for $\alpha\rightarrow 1$, as soon as
$\alpha<1$, a strong variation in $t_2/t_1$ is observed
for large $E_J/E_C$.
In Fig.~\ref{fig:t21vsTau}, we plot $t_2/t_1$ versus $E_J/E_C$ for
different  values of $\alpha$. 
For $\alpha=1$, the curve is almost a constant. 
In Table \ref{tab}, we report a set of quantities calculated by varying
both $\alpha$ and $E_J/E_C$, such as to keep the energy gap $\Delta_0$ at
zero applied voltage of the  order of $\approx 0.1 E_C$.
\begin{figure}[t]
   \begin{center}
     \includegraphics[width=8.25cm,keepaspectratio]{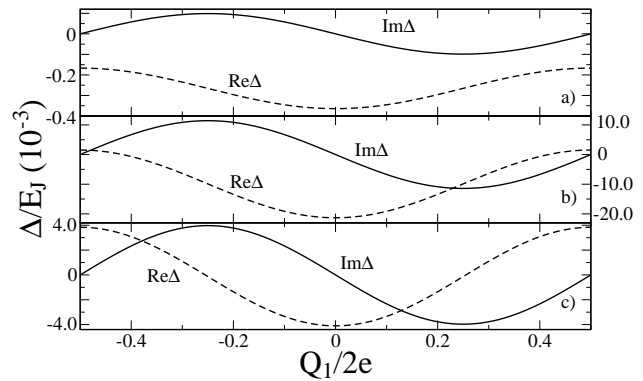}
     \caption{\small Plot of the real and imaginary part of
                     $\Delta$  as a function of $Q_1/2e=C V_1/2e$ 
                     for $C V_2/2e=0.5$ for 
                     a) $\alpha=0.95$, $E_J/E_C=35$;
                     b) $\alpha=0.95$, $E_J/E_C=10$; and  
		     c) $\alpha=1$, $E_J/E_C=15$. 
                     \label{fig:ReImQ1}}
   \end{center}
\end{figure}

\begin{table}[b]
\begin{tabular}{cc|ccc ccc}\hline\hline
$\alpha$ & $E_J/E_C$ & $t_2/t_1$ & \ $t_1/E_J$  &  $t_2/E_J$  &
$\frac{|\Delta|_0}{E_J}$  &
  $\frac{|\Delta|_{\textrm{min}}}{|\Delta|_0}$ 
& $\frac{E_{12}}{|\Delta|_0}$ \\ 
 & & & $\times10^{-3}$ & $\times 10^{-5}$ &  &  & \\
\hline 
0.80 & 35 & 0.0062 &  -2.9   &  -1.8 & 0.0059 & 0.98 & 82\\
0.85 & 30 & 0.030  &  -1.9   &  -5.8 & 0.0040 & 0.88 & 126\\ 
0.90 & 25 & 0.12   &  -1.5   &  -18  & 0.0037 & 0.61 & 149\\ 
0.95 & 20 & 0.39   &  -1.5   &  -59  & 0.0054 & 0.12 & 116\\ 
1.00 & 15 & 0.97   &  -2.05  & -198  & 0.012  & 0    & 61\\ 
1.05 & 10 & 1.77   &  -4.2   &  -740 & 0.038  & 0    & 24\\ 
\hline\hline
\end{tabular}
\caption{Values of $t_1$, $t_2$, their
  ratio $t_2/t_1$, the energy gap $|\Delta|_0$ at zero applied voltage, and the 
  minimum of the gap $|\Delta|_{\rm min}$ 
  for a series of values
  of $\alpha$ and $E_J/E_C$. In the last column we report the ratio of
  the energy difference $E_{12}$ between the second and first excited state and
  the qubit gap $|\Delta|_0$.\label{tab}} 
\end{table}

The parameters of an experimentally realized flux qubit (Delft qubit) 
\cite{Chiorescu03} are $\alpha=0.8$ and $E_J/E_C=35$ and are
given in the first row of Table~\ref{tab}.
In this case, the ratio $t_2/t_1$ is very
small and the contribution of  $t_2$ is negligible.
This choice of parameters of the Delft qubit therefore does not allow the 
manifestation of a significant $\sigma_y$ term in the single-qubit 
Hamiltonian, for any value of the bias voltage.

\begin{figure}[t]
   \begin{center}
     \includegraphics[width=8cm,keepaspectratio]{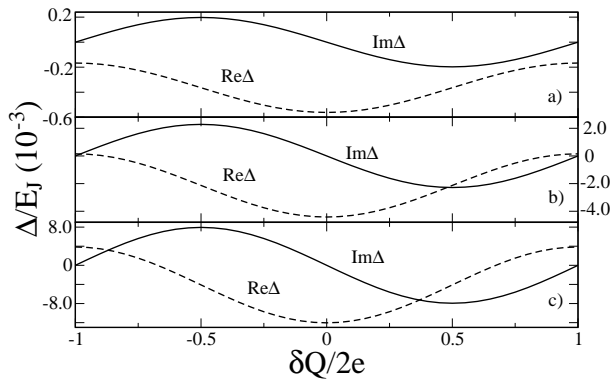}
     \caption{\small Plot of the real and imaginary part of
                     $\Delta$  as a function of $\delta
                     Q/2e=C(V_1-V_2)/2e$ for $V_1+V_2=0$ 
                     choosing a) $\alpha=0.95$,
                     $E_J/E_C=35$, b) $\alpha=0.95$, $E_J/E_C=10$ and  
		     c) $\alpha=1$, $E_J/E_C=15$. 
		     \label{fig:ReIm}}
   \end{center}
\end{figure}

\begin{figure}[t]
   \begin{center}
     \includegraphics[width=8cm,keepaspectratio]{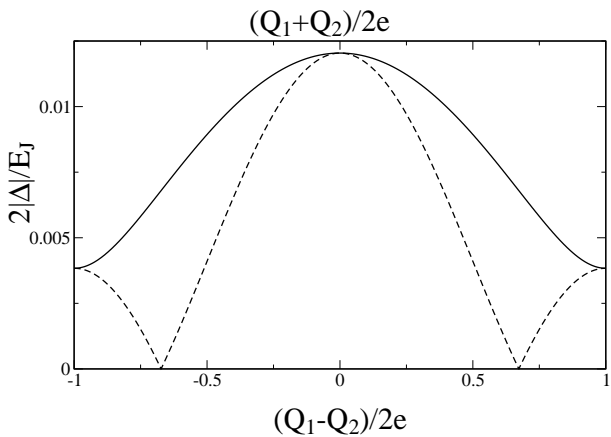}
     \caption{\small Plot of the gap versus
                     $\delta Q/2e=\gamma C_J(V_1-V_2)/2e$ (solid line)
                     and $\gamma C_J(V_1+V_2)/2e$  (dashed line), for
                     $\alpha=1$ and $E_J/E_C=15$. In this case both
                     the amplitude of oscillation and the cross region
                     of the curves are appreciable.\label{fig:delta-gap}}
   \end{center}
\end{figure}

In Fig.~\ref{fig:ReImQ1}, we plot the real and imaginary part
of $\Delta$ as a function of the applied voltage $V_1$, expressed
in the gate charge $Q_1=C_1 V_1$, while keeping the other gate
voltage fixed such that $Q_2/2e=C_2 V_2/2e = 0.5$.
If the real part of $\Delta$ can be tuned from a finite value
to zero while the imaginary part of $\Delta$ remains finite
(as in Fig.~\ref{fig:ReImQ1}c),
then the pseudo-field ${\bf B}$ can point along arbitrary angles in the
equator plane of the Bloch sphere.
The magnitude of the pseudo-field can be controlled in principle
by changing $\alpha$, e.g., with a flux-tunable junction.
In Fig.~\ref{fig:ReIm}, we plot the real and imaginary part of
$\Delta$ in the case where both voltages are varied simultaneously
such that $V_1=-V_2$ as a function of   
$\delta Q/2e=C(V_1-V_2)/2e$.  
In Fig.~\ref{fig:delta-gap} we plot the gap $|\Delta|$ as a function
of $\delta Q/2e=C(V_1-V_2)/2e$ (solid line) and of
$(Q_1+Q_2)/2e=C(V_1+V_2)/2e$ (dashed line) for this set of
parameters. 

\section{Full control for $\alpha>1$}
\label{sec:full-control}

\begin{figure}[b]
   \begin{center}
     \includegraphics[width=8cm]{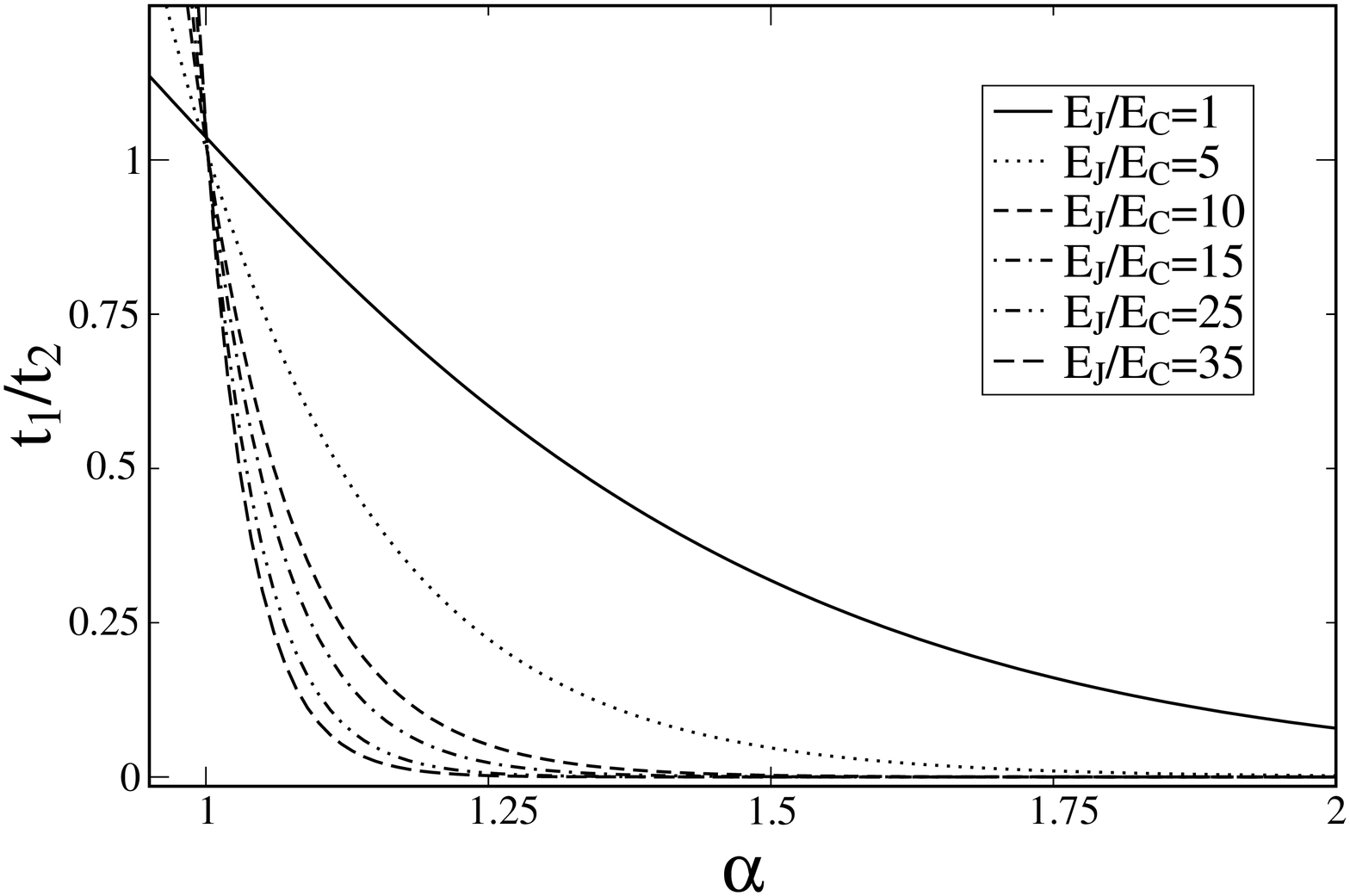}
     \caption{\small The ratio 
             $t_1/t_2$ between the tunneling matrix elements, 
             plotted as a function of
             $\alpha\ge 1$ for  several values of
             $E_J/E_C$.\label{fig:t12-vs-a}}
   \end{center}
\end{figure}

\begin{figure}[b]
   \begin{center} 
     \includegraphics[width=8cm]{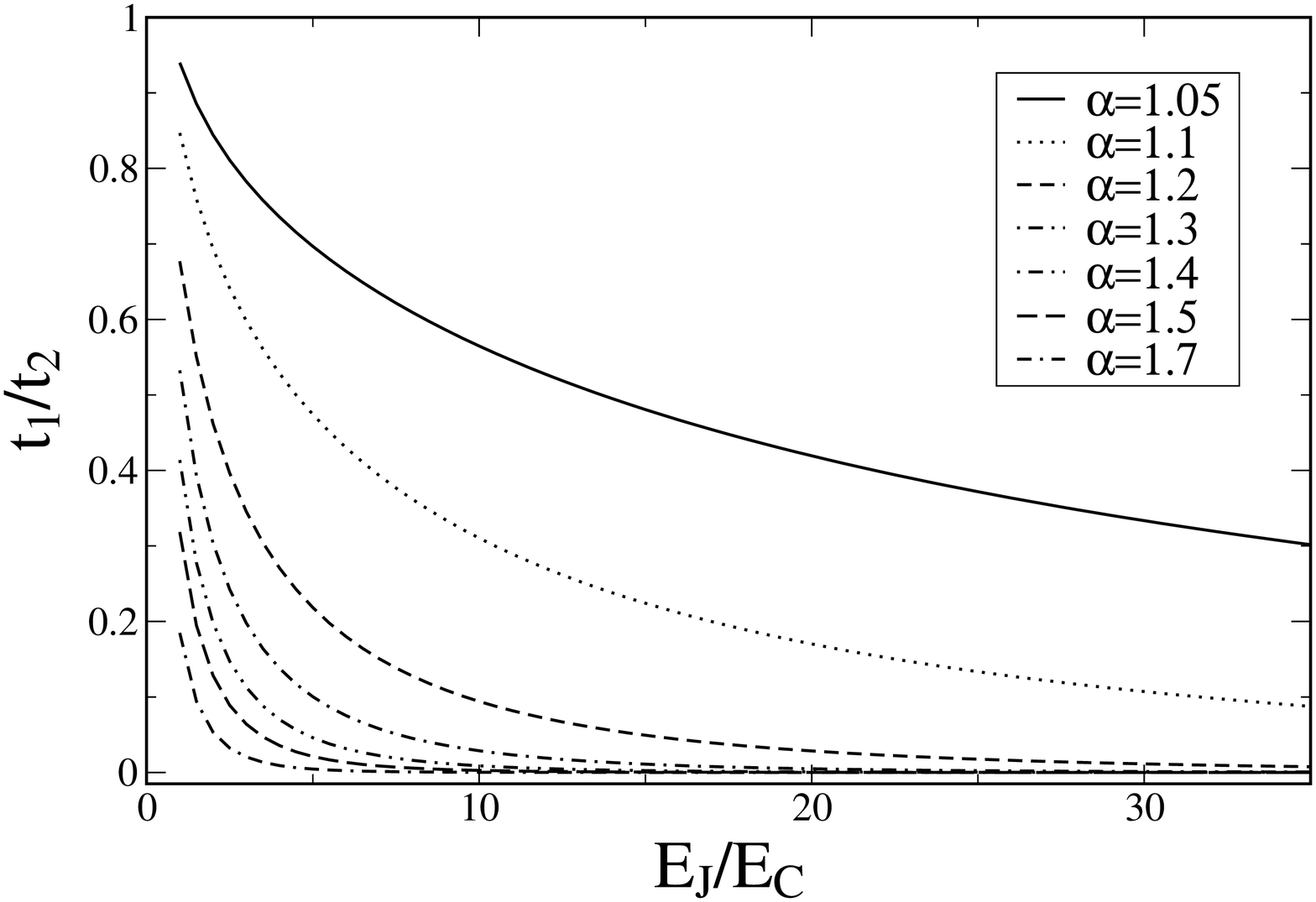}
     \caption{\small The ratio 
             $t_1/t_2$ between the tunneling matrix elements,
             plotted as a function of
             $E_J/E_C$  for several values of
             $\alpha\ge 1$.\label{fig:t12-vs-tau}}
   \end{center}
\end{figure}

The flux qubit realized at Delft \cite{Chiorescu03} operates with a ratio 
$\alpha=0.8<1$ between the Josephson energies of its junctions.
As shown in Table~\ref{tab}, the ratio of tunneling matrix elements
for this parameter choice is $t_2/t_1=0.0062$, thus the effect of the
applied voltages is negligible. 
Two other regimes for $\alpha$ are interesting, namely
$\alpha\approx1$ and $\alpha>1$. 

In the former, $t_1$ and $t_2$ are
approximately equal.
In this case, $\bphi$ can tunnel from a
left minimum (L) to a right one (R) via both an intra-cell or an
inter-cell tunneling process with almost equal probability.
However, while inter-cell
tunneling can be controlled via the applied voltages $V_1$ and $V_2$, 
allowing superposition
with non-zero relative phase of the qubit states, the intra-cell
transition amplitude remains constant, once the parameters $\alpha$ and
$E_J/E_C$ are fixed, thus leading only to qubit flips. 
In Table~\ref{tab}, for each value of $\alpha<1$, the minimum of the gap  
is a finite quantity and can be calculated by minimization of equation  
Eq.~(\ref{eigenenergy}) with respect to ${\bf k}$.
However, for $\alpha\ge1$ there is  a value of the external applied
voltage for which the gap goes to zero (Fig.~\ref{fig:delta-gap}).

\begin{figure}[b]
   \begin{center}
     \includegraphics[width=8.0cm]{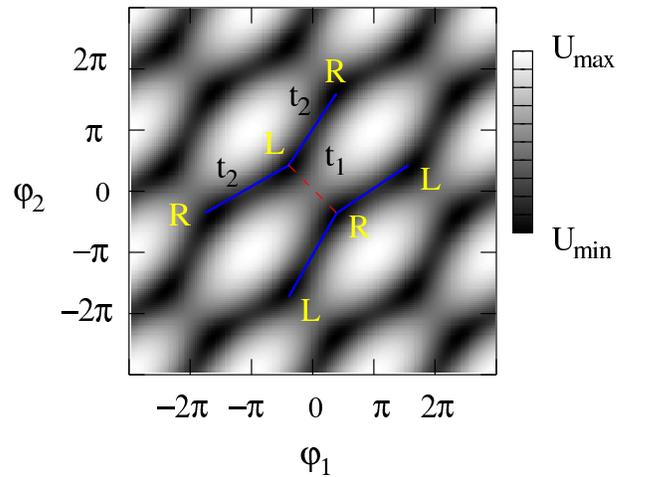}
     \caption{Density plot of the double well potential
              $U(\varphi_1,\varphi_2)$ for
              $\alpha=1.4$, on a logarithmic scale. Two
              equivalent one-dimensional chains  with nearest
              neighbor interaction are highlighted in the figure.   
	      \label{fig:1D-chain}}
   \end{center}
\end{figure}

We are particularly interested in the regime $\alpha>1$. In this case
$t_1<t_2$, i.e., the intra-cell tunneling between two minima is inhibited
 and, with a suitable choice of $\alpha$, can be completely suppressed
(Figs.~\ref{fig:t12-vs-a} and \ref{fig:t12-vs-tau}).
In this situation, the system can be described by a one-dimensional
chain in which every even (odd) site is labeled as a ``left'' minimum L
while the remaining sites are labeled ``right'' minima R, see
Fig.~\ref{fig:1D-chain}.
The tunneling matrix element between the sites is $t_2$ ($t_1=0$).
Note that, due to the periodicity of the system, all L (R) sites have
to be identified with each other, since they describe the same
configuration.

  From Eqs.~(\ref{ReD}) and (\ref{ImD}), we immediately find that,
for $t_1/t_2\rightarrow 0$, we gain full control of the direction 
of the effective pseudo-field ${\bf B}$ in
the equatorial plane of the Bloch sphere, since
\begin{equation}
\label{Delta2}
\Delta(k_+,k_-)=4t_2\cos(\pi k_+)e^{i\pi k_-},
\end{equation}
where $k_\pm=(C_1 V_1 \pm C_2 V_2)/2e$.
The sum and difference of the gate charges therefore independently
control the qubit energy gap and the angle $\theta$ of the 
pseudo-field,
\begin{equation}
|\Delta| = 4|t_2\cos(\pi k_+)|,\quad\quad
\theta = \pi k_-.\label{eq:gap}
\end{equation}

\section{Charge decoherence}
\label{sec:decoherence}

Voltage fluctuations from imperfect voltage sources or
other fluctuating charges in the environment lead to
charge fluctuations on the two islands in the circuit
and thus to decoherence of the qubit.
Moreover, we are considering here a situation where the
sensitivity to external voltages has been deliberately enhanced
and therefore it can be expected that charge fluctuations
cannot be ignored.
An estimate of the decoherence time for the same circuit has been
developed in \cite{LinTian99}, where it is found to be $0.1\,{\rm s}$.

In order to model bias voltage fluctuations, we include the
two impedances $Z_1$ and $Z_2$ (Fig.~\ref{fig:circuit}) in
our analysis.
  From circuit theory \cite{charge}, we can then obtain a 
Caldeira-Leggett model for the system coupled to its
charge environment,
\begin{equation}
  \label{eq:CL}
  {\cal H} = {\cal H}_S + {\cal H}_B + {\cal H}_{SB},
\end{equation}
where ${\cal H}_S$ from Eq.~(\ref{H}) describes the dissipationless
elements of the circuit, and
\begin{equation}
  \label{eq:HB}
  {\cal H}_B = \sum_{j=1,2}\sum_\nu \left(\frac{p_{j\nu}^2}{2m_{j\nu}}
 + \frac{1}{2}m_{j\nu}\omega_{j\nu}^2x_{j\nu}^2 \right),
\end{equation}
is the Hamiltonian of the degrees of freedom of two independent
baths of harmonic oscillators that are used to model the two
impedances, and finally
\begin{equation}
  \label{eq:HSB}
  {\cal H}_{SB} = \sum_{j=1,2} {\bf m}_j \cdot {\bf Q}
  \sum_\nu c_{j\nu} x_{j\nu},
\end{equation}
describes the system-bath coupling,
where ${\bf m}_1={\cal C}^{-1}(C_1 , 0)^T$ 
and ${\bf m}_2={\cal C}^{-1} (0 ,  C_2)^T$.
The coupling constants $c_{j\nu}$ are related to $Z_j$
via the spectral densities
\begin{equation}
  \label{eq:HSBc}
  J_j(\omega) = -\omega{\rm Re}Z_j(\omega) 
    = \frac{\pi}{2}\sum_\nu \frac{c_{j\nu}^2}{m_{j\nu}\omega_{j\nu}}\delta(\omega-\omega_{j\nu}).
\end{equation}
The decoherence rates in the Born-Markov approximation are given by
\cite{charge}
\begin{eqnarray}
  \frac{1}{T_1} \!&=& \!\frac{4}{\hbar^2}\sum_{j=1,2}
    \left|{\bf m}_j\cdot\langle 0|{\bf Q}|1\rangle\right|^2
  \Delta{\rm Re Z}_j(\Delta)\coth\frac{\Delta}{2k_B
    T},\label{eq:T1}\nonumber\\ 
  \\
  \frac{1}{T_\phi}\! &=& \!\frac{1}{\hbar^2}\sum_{j=1,2}
  \left|{\bf m}_j\cdot\left(\langle 0|{\bf Q}|0\rangle\!-\!\langle 1|{\bf Q}|1
  \rangle\right)\right|^2 {\rm Re Z}_j(0) 2k_B T.\label{eq:Tphi} \nonumber\\ 
\end{eqnarray}

Now we compute the matrix elements of the charge
operator ${\bf Q}=-2ie\bnabla$ in the
$|0\rangle$, $|1\rangle$ basis. Following the derivation of
the Hamiltonian in Sec.~\ref{ssec:tight-binding}, we start from 
\begin{equation}\label{Q-mat-el}
\langle u_{\alpha{\bf k}}|{\bf Q}|u_{\beta{\bf k}}\rangle=-2e{\bf
  k}\delta_{\alpha\beta}-2ie\langle\psi_{\alpha{\bf k}}|\bnabla
  |\psi_{\beta{\bf k}}\rangle.
\end{equation}
The matrix elements of ${\bf Q}$ between the Bloch states 
\begin{equation}
\langle \psi_{\alpha{\bf k}}|{\bf Q}|\psi_{\beta{\bf k}}\rangle=\sum_{{\bf n},{\bf
    m}\in{\bb Z}^2}\e^{2\pi i {\bf k}\cdot({\bf n}-{\bf
    m})}{\bf Q}_{\alpha{\bf m},\beta{\bf n}},
\end{equation}
are given in terms of the matrix elements of $\bnabla$ between
the Wannier functions
\begin{eqnarray}
{\bf Q}_{\alpha{\bf n},\beta{\bf m}}&=&
-2ei\langle\phi_{\alpha{\bf n}}|\bnabla|\phi_{\beta{\bf m}}\rangle\nn\\
&=&-2ei\left({\cal G}^T{\bf P}{\cal G}\right)_{\alpha,{\bf n},\beta{\bf m}},
\end{eqnarray}
and, in turn, through the ${\cal G}$-matrix, they are expressed in terms 
of the Gaussian states, 
\begin{equation}
{\bf P}_{\alpha{\bf n},\beta{\bf m}} = 
\langle \Psi_{\alpha{\bf n}}|\bnabla|\Psi_{\beta{\bf m}}\rangle
= \frac{1}{2}{\cal M}\Delta\bphi_{\alpha{\bf n},\beta{\bf
    m}}S_{\alpha{\bf n},\beta{\bf m}},
\end{equation}
where the matrix ${\cal M}$ is defined in Eq.~(\ref{M}),
$\Delta\bphi_{\alpha{\bf n},\beta{\bf
    m}}=\bphi_{\beta}-\bphi_{\alpha}+2\pi({\bf m}-{\bf n})$, and the
$S$-matrix is defined in Eq.~(\ref{S-matrix}).

We only keep the leading matrix elements $s_1$ and $s_2$
 in the overlap matrix $S$ when calculating
the ${\cal G}$ and  ${\bf P}$ matrices (see
Sec.~\ref{sec:t1t2}). 
Since the largest contributions
of ${\bf P}$ are proportional to $s_1$ and $s_2$, we can 
use ${\cal G}\approx\openone$, and thus 
${\bf Q}_{\alpha{\bf n},\beta{\bf m}}\simeq
{\bf P}_{\alpha{\bf n},\beta{\bf m}}
\propto {\bf S}_{\alpha{\bf n},\beta{\bf m}}$.
We consider the diagonal term and the
off-diagonal term separately and obtain,
\begin{eqnarray}
\langle u_{\alpha{\bf k}}|{\bf Q}|u_{\alpha{\bf k}}\rangle
&=&-{\bf Q}_g,\label{Qdiag}\\
\langle u_{{\rm L}{\bf k}}|{\bf Q}|u_{{\rm R}{\bf k}}\rangle &=&
-ei{\cal M}\Big[s_1 \Delta\bphi 
+s_2(\Delta\bphi-2\pi{\bf  e}_1)\e^{2\pi ik_1}\nn\\
& & + s_2(\Delta\bphi+2\pi{\bf e}_2)\e^{-2\pi ik_2}\Big],\label{Qoff}
\end{eqnarray}
where $s_1$, $s_2$,
$\Delta\bphi=\bphi_{\rm R}-\bphi_{\rm L}$, 
and the matrix ${\cal M}$ depend
on $\alpha=E_{J3}/E_J$ and $E_J/E_C$. 
In the qubit basis we find,
\begin{eqnarray}
\langle 0|{\bf Q}|0\rangle&-&\langle 1|{\bf Q}|1\rangle=
-e{\cal M}\Big[s_1\sin(\theta)\Delta\bphi\nn\\
&+&s_2\sin(\theta+2\pi k_1)(\Delta\bphi-2\pi{\bf
  e}_1)\label{eq:Q-mat-el}\nn\\
&+&s_2\sin(\theta-2\pi k_2)(\Delta\bphi+2\pi{\bf
  e}_2)\Big],\label{eq:Q-00-11}\\
\langle 0|{\bf Q}|1\rangle&=&ie{\cal
  M}\Big[s_1\cos(\theta)\Delta\bphi\nn\\
&+&s_2\cos(\theta+2\pi
  k_1)(\Delta\bphi-2\pi{\bf e}_1)\nn\\
&+&s_2\cos(\theta-2\pi
  k_2)(\Delta\bphi+2\pi{\bf e}_2)\Big],\label{eq:Q-01}
\end{eqnarray}
where $\tan\theta={\rm Im}\Delta/{\rm Re}\Delta$ is a function of
$k_{1,2}=C_{1,2}V_{1,2}/2e$.
Using Eqs.~(\ref{eq:T1}), (\ref{eq:Tphi}), (\ref{eq:Q-00-11}), and
(\ref{eq:Q-01}) we can express the decoherence rates in a more 
explicit way,
\begin{eqnarray}
\frac{1}{T_1}&=& 2\pi\frac{E_J}{\hbar}\frac{{\rm Re}Z}{R_Q}
\left(\frac{C}{C_J}\right)^2s_2^2~{\cal F}_1(V_1,V_2),\label{eq:T1-pref}\\
\frac{1}{T_{\phi}}&=&2\pi\frac{2k_BT}{\hbar}\frac{{\rm Re}Z}{R_Q}
\left(\frac{C}{C_J}\right)^2s_2^2~{\cal
  F}_{\phi}(V_1,V_2),\label{eq:Tphi-pref}
\end{eqnarray}
where $s_2$, ${\cal F}_1$, and ${\cal F}_{\phi}$ are given in the
Appendix~\ref{sec:s1-s2}. ${\cal F}_1$ and ${\cal F}_{\phi}$ are
periodic functions of the applied voltages $V_1$ and $V_2$ that depend
on the parameters $\alpha$, $E_J/E_C$, and on $s_1/s_2$. They can be
estimated to be at most of order one, depending on the choice of
parameters and the applied voltages. 
In Eqs.~(\ref{eq:T1-pref}) and (\ref{eq:Tphi-pref}) we
chose $Z\approx Z_1\approx Z_2$,  and $R_Q=h/e^2$ is the quantum of
resistance.  

In the regime $\alpha>1$ we have $s_2\gg s_1$. For $\alpha=1.4$,
$E_J/E_C=15$ and $C/C_J=0.02$ we find that $s_2=8 \cdot 10^{-4}$. An
estimate for
$T\approx 100\,{\rm mK}$, ${\rm Re}Z\approx 1\,{\rm k}\Omega$ 
and $E_J=250 \, {\rm GHz}$ produces decoherence times in the
millisecond range,
\begin{eqnarray}
\frac{1}{T_1}&\simeq&  \frac{{\cal F}_1(V_1,V_2)}{{\cal F}_{1,{\rm
    max}}} \frac{1}{6\,{\rm ms}}, \label{eq:T1-a>1}\\ 
\frac{1}{T_{\phi}}&\simeq& \frac{{\cal F}_{\phi}(V_1,V_2)}{{\cal
  F}_{\phi,{\rm max}}}  \frac{1}{12\,{\rm ms}}.  \label{eq:Tphi-a>1}
\end{eqnarray} 
For some particular values of $V_1$ and $V_2$ the functions
${\cal F}_1$ or ${\cal F}_{\phi}$ vanish, implying that
$1/T_1\rightarrow 0$ or $1/T_{\phi} \rightarrow 0$. 
In particular, ${\cal F}_1=0$ for $(C_1 V_1,C_2 V_2)/2e
=\pm(1/2,0),~ \pm(0,1/2),
~\pm(1/4,1/4),~\pm(1/8,-1/8),~\pm(3/8,-3/8)$ in the FBZ, 
and ${\cal F}_{\phi}=0$ for 
$(C_1 V_1,C_2 V_2)/2e=(n/2,m/2),~\pm(1/4,-1/4)+(n,m)$,
with $n,m\in{\bb Z}$. The two functions have a
common set of zeros, namely $\pm(n/2,0), \pm(0,m/2)$, with
$n,m\in{\bb Z}$. In these cases, both $1/T_1, 1/T_{\phi}\rightarrow 0$.

For the regime $\alpha<1$ we have that $s_1\gg s_2$ and we can neglect
terms containing $s_2$. Choosing $\alpha=0.8$ and $E_J/E_C=35$ we find
$s_1=1.3\cdot 10^{-5}$. It follows that the decoherence rates are
strongly suppressed and an estimate shows that they are below $1\,
{\rm Hz}$.  This means that in this case the main process that causes
decoherence is not due to the charge degrees of freedom.  In fact for
the Delft qubit \cite{Chiorescu03}, that operates in this regime, the
dephasing and the relaxation times caused by other mechanisms are much
smaller, $T_{\phi}=20~{\rm ns}$ and $T_1=900\,{\rm ns}$.   

The physical reason for the small decoherence and relaxation rates
found here is that, despite the voltage bias, we are still dealing 
with a flux qubit whose states are indistinguishable from their 
charge distribution, as seen from Eq.~(\ref{Qdiag}).

\section{Results and conclusions}
\label{sec:results}

By means of circuit theory and a tight-binding approximation, we
have analyzed a voltage-controlled SC flux qubit circuit that allows
full control of the single-qubit Hamiltonian Eq.~(\ref{qubitH}), 
with $\sigma_x$, $\sigma_y$ and $\sigma_z$ terms, in order to allow 
arbitrary single qubit operations. 

One of the main results of this work is the computation of the tunneling
matrix elements appearing in the single qubit Hamiltonian as a
function of the device parameters $\alpha$ and $E_J/E_C$. This allowed
us to explore new possible working regimes of the system, looking for
a range of parameters for which a full control on qubit rotations is
feasible.
Substantially, the qubit can work in two different regimes,
$\alpha<1$ and $\alpha>1$, showing different features. In particular,
for $\alpha>1$, the pseudo magnetic field ${\bf B}$ that couples to
the qubit in the Hamiltonian has a non-zero $y$-component. This allows
full control of qubit rotations on the Bloch sphere through the
applied voltages $V_1$ and $V_2$.
In fact, in the Hamiltonian, Eq.~(\ref{qubitH}), the off-diagonal term
$\Delta$, given in Eq.~(\ref{delta}), contains the voltages $V_{1,2}$ and
the sensitivity to $V_{1,2}$ is determined by the tunneling parameters
$t_1$ and $t_2$ in Eqs.~(\ref{t1}), (\ref{t2}).

For $\alpha\le 1$, we find $t_1\gtrsim t_2$. The effect of $t_2$, and
thus of the applied voltages, for the value of parameters of the
Delft qubit \cite{Chiorescu03}, is negligible as shown in
Table \ref{tab}, but can be greatly enhanced for a suitable
choice of $\alpha$ and $E_J/E_C$ (see Figs. \ref{fig:t21vsAlpha} and
\ref{fig:t21vsTau}), thus allowing good control in the real
and imaginary parts of $\Delta$, as shown in Eqs.~(\ref{ReD}) 
and (\ref{ImD}) and in Figs.~\ref{fig:ReImQ1} and \ref{fig:ReIm}. 

In the case $\alpha>1$, the roles of $t_1$ and $t_2$ are interchanged,
as shown in Figs.~\ref{fig:t12-vs-a}, \ref{fig:t12-vs-tau}, and a new
regime in which a full control of the single-qubit Hamiltonian
becomes possible. 
For a suitable choice of $\alpha$ and $E_J/E_C$, the tunneling
parameter $t_1$ become vanishingly small, 
giving rise to a simple dependence of $\Delta$
on the voltages, as found in Eqs.~(\ref{Delta2}) and (\ref{eq:gap}).

Our analysis is based on the two-level approximation, i.e., we assume
that we can neglect all high levels besides the two lowest ones. 
This approximation is justified if the energy gap $E_{12}$
between the two lowest levels and any higher level is
sufficiently large, in particular, larger than the qubit gap
$E_{01}=|\Delta|$. The gap $E_{12}$ can be roughly estimated as the
plasma frequency, i.e., the smallest of the frequencies of the
(anisotropic) harmonic oscillator arising from the linearization of
the equation of motion around the minimum configurations of the
potential. This frequency is given by (also see Appendix \ref{sec:B})
$\omega_{LC}=1/\sqrt{C_JL_J}=\sqrt{8E_JE_C}/\hbar$. 
In Table~\ref{tab}, we report the ratio of $E_{12}$ 
and the qubit gap $|\Delta_0|$ at zero applied voltage.
For all parameter values studied, $E_{12}$ exceeds
$2|\Delta_0|$ by more than a factor of 20, in many relevant 
cases even by two orders of magnitude, thus justifying the
two-level approximation. 

Finally, we have studied the decoherence due to
charge fluctuations of the voltage sources.
Our result for the $T_1^{-1}$ and $T_{\phi}^{-1}$ rates is given
in Eqs.~(\ref{eq:T1-pref}) and (\ref{eq:Tphi-pref}), 
an estimate of which yields a coherence time
longer than $\approx 1\,{\rm ms}$, leading to the conclusion that charge
fluctuations are not the main source of decoherence, even in the
regime in which the sensitivity to external voltages is enhanced. 
The coherence of the system is well preserved, since the qubit is
still essentially a SC flux qubit, i.e., the $|0\rangle$ and
$|1\rangle$ states have nearly identical charge configurations.

In conclusion, based on our analysis we find that full control of
single-qubit operations in a SC flux qubit should be feasible, 
provided that the right choice of the device parameters is made.

\section*{Acknowledgments}
We would like to thank David DiVincenzo for very useful
discussions.
We acknowledge financial support from the 
Swiss National Science Foundation.

\appendix

\section{Matrices ${\cal C}$, ${\bf C}_V$ , ${\bf M}_0$, and ${\bf N}$}
\label{sec:A}

The definitions of the derived matrices ${\cal C}$, ${\bf C}_V$, ${\bf M}_0$ 
and ${\bf N}$ that enters the Hamiltonian are given in
\cite{charge,BKD} for the general case. Here we apply the theory and
derive the matrices for the particular case of the circuit of
Fig.~\ref{fig:circuit}. The derived capacitance matrices are
\begin{eqnarray}
{\cal C}&\equiv&{\bf C}_J+\left(\begin{array}{cc}
{\bf C} & 0\\
0 & 0\end{array}\right),\\
{\bf C}_V&\equiv&({\bf C},{\bf 0})^T.
\end{eqnarray}
The inductance matrices that enter the potential are  
\begin{eqnarray}
{\bf M}_0&=&\frac{1}{K}{\bf F}_{JK}{\bf F}_{JK}^T,\\
{\bf N}&=&-\frac{1}{K}{\bf F}_{JK},
\end{eqnarray}
and ${\bf M}_0^T={\bf M}_0$.
For the circuit studied here, we obtain
\begin{eqnarray}
  {\bf M}_0 \!\! &=& \!\!\frac{1}{K}\left(\begin{array}{ccc} 1 & -1 & -1\\ -1 &
    1 & 1\\ -1 & 1 & 1\end{array}\right), \,\,
  {\bf N} =
    \frac{1}{K}\left(\begin{array}{c} 1\\-1\\-1\end{array}\right)\!.
\end{eqnarray}

\section{Projected matrices}
\label{sec:B}

The three-dimensional problem is mapped into a
two-dimensional one in Sec.~\ref{sec:born-oppenheimer} with
the matrix
\begin{equation}
{\cal P}=\left(\begin{array}{cc} 1 & 0\\ 0 & 1\\ 1 &
-1\end{array}\right),
\end{equation}
via the relation 
$(\varphi_1,\varphi_2,\varphi_3)^T={\cal P}(\varphi_1,\varphi_2)^T$.
In the case of symmetric double well potential, the inductance
linearized matrix ${\bf L}^{-1}_{\rm lin;L,R}$ is given by
\begin{equation}
{\bf L}_{\textrm{\rm lin;L,R}}^{-1}={\bf M}_0
+{\bf L}_{J}^{-1}\boldsymbol{\cos}\bphi_{\rm L,R;i}.
\end{equation}
Because of the symmetry of the potential, we drop 
the subscripts R and L.
Applying the matrix ${\cal P}$ we obtain  
${\bf L}^{-1}_{{\rm lin},P}={\cal P}^T {\bf L}_{\rm lin}^{-1}{\cal P}$,
\begin{equation}
{\bf L}^{-1}_{{\rm lin},P}=\frac{1}{L_J}\left(\begin{array}{cc} 
\alpha & \frac{1}{2\alpha}-\alpha\\ 
\frac{1}{2\alpha}-\alpha & \alpha\end{array}\right).
\end{equation}
 In order to simplify the calculation we assume the two
capacitance $C_1$ and $C_2$ to be equal, $C_1=C_2\equiv C$
and define $\gamma=C/C_J$.
The projected capacitance matrix  ${\cal C}_P={\cal P}^T{\cal C}{\cal P}$
is then found to be
\begin{equation}
{\cal C}_P=C_J\left(\begin{array}{cc} 1+\gamma+\alpha & -\alpha\\
-\alpha & 1+\gamma+\alpha\end{array}\right).
\end{equation}
In this case, the orthogonal matrices that diagonalize the capacitance  
matrix ${\cal C}_P$ the  linearized 
inductance matrix ${\bf L}^{-1}_{{\rm lin},P}$ are identical,   
${\cal C}_P={\bf O}^T{\cal C}_d{\bf O}$ and  ${\bf L}^{-1}_{{\rm lin},P}=
{\bf O}^T{\bf \Lambda}{\bf O}$.
The frequency matrix 
${\bf \Omega}=\textrm{diag}(\omega_{\perp},\omega_{\parallel})$
is given by
\begin{equation}
{\bf \Omega}^2=\omega_{LC}^2
\left(\begin{array}{cc}
\frac{1}{4\alpha^2(1+\gamma)^2} & 0\\
0 & \frac{1-4\alpha^2}{4\alpha^2(1+2\alpha+\gamma)^2}
\end{array}\right),
\end{equation}
where $\omega_{LC}^2=1/L_J C_J$.
The matrix ${\cal M}$ is then diagonalized by the same orthogonal
matrix ${\bf O}$ and, in the basis where it is diagonal, can be
written as 
\begin{equation}
{\cal M}=\sqrt{\frac{E_J}{8E_C}}
\left(\begin{array}{cc}
\sqrt{\frac{1+\gamma}{2\alpha}} & 0\\
0 & \sqrt{\frac{(4\alpha^2-1)(1+2\alpha+\gamma)}{2\alpha}}
\end{array}\right).
\end{equation}

\section{The functions ${\cal F}_1$ and ${\cal F}_{\phi}$}
\label{sec:s1-s2}

We give here an explicit formula for the intra-cell and inter-cell
overlaps $s_1$ and $s_2$ as functions of $\alpha$, $E_J/E_C$ and $C/C_J$,
\begin{widetext}
\begin{eqnarray}
s_1&=&\exp\left\{-\frac{E_J}{4\sqrt{2\alpha}E_C}\arccos^2\left(\frac{1}{2\alpha}\right)
\sqrt{(4\alpha^2-1)(1+2\alpha+C/C_J)}\right\},\label{eq:s1}\\
s_2&=&\exp\left\{-\frac{E_J}{16E_C}
  \left[\pi^2\sqrt{\frac{1+C/C_J}{2\alpha}}+\left(\pi-2\arccos 
\left(\frac{1}{2\alpha}\right)\right)^2\sqrt{\frac{(4\alpha^2-1)(1+2\alpha+C/C_J)}{2\alpha}}
  \right]\right\}.\label{eq:s2}    
\end{eqnarray} 
Through these quantities we can express ${\cal F}_1$ and ${\cal F}_{\phi}$ as functions of
$k_1$ and $k_2$, with $k_i=C_iV_i/2e$,
\begin{eqnarray}
{\cal F}_1(k_1,k_2)&=&\frac{|\Delta(k_1,k_2)|}{E_J}\coth
\left(\frac{|\Delta(k_1,k_2)|}{2k_BT}\right)\tilde{\cal F}_{\phi}(k_1,k_2),\\
{\cal F}_{\phi}(k_1,k_2)&=&\frac{4}{\det^2({\cal C})}\sum_{i=1,2}\left[\pi ({\cal
    C}_{1i}{\cal M}_{22}\sin(2\pi k_2-\theta)-{\cal C}_{2i}{\cal
    M}_{11}\sin(2\pi k_1+\theta))\right.\nn\\
&+&\left.({\cal C}_{2i}{\cal M}_{11}+{\cal C}_{1i}{\cal M}_{22}) 
\arccos\left(\frac{1}{2\alpha}\right)\left(\frac{s_1}{s_2}\sin(\theta)+\sin(2\pi
  k_1+\theta) -\sin(2k_2-\theta)\right)\right]^2\label{eq:F},
\end{eqnarray}
where $\tilde{\cal F}_{\phi}$ is given by ${\cal F}_{\phi}$, once the
$\sin$ are replaced by $\cos$. ${\cal C}_{ij}$ and ${\cal M}_{ij}$ are
the entries of the matrices ${\cal C}$ and ${\cal M}$ defined in
Appendix~\ref{sec:B}. The gap $|\Delta|$ and the relative phase 
between the states $|0\rangle$ and $|1\rangle$ are given by
\begin{eqnarray}
|\Delta(k_1,k_2)|&=&2\sqrt{(t_1+2t_2\cos[\pi(k_1-k_2)]\cos[\pi(k_1+k_2)])^2
+4t_2^2\cos[\pi(k_1+k_2)]\sin[\pi(k_1-k_2)]},\\
\tan\theta&=&\frac{2t_2\cos[\pi(k_1+k_2)]\sin[\pi(k_1-k_2)]}{t_1+
  2t_2\cos[\pi(k_1-k_2)]\cos[\pi(k_1+k_2)]}. 
\end{eqnarray}
\end{widetext}

\end{document}